\documentclass[letterpaper]{article}
\pdfoutput=1
\usepackage{jheppub}
\usepackage{amsmath}
\usepackage{graphicx}
\usepackage{array}
\usepackage{slashed}
\usepackage{appendix}
\usepackage{multirow}
\usepackage{xcolor}
\usepackage{aas_macros} 

\newcommand{\bea}{\begin{eqnarray}}
\newcommand{\eea}{\end{eqnarray}}

\newcommand{\anomaly}{\mathcal{A}}

\newcommand{\xperp}{\vec{x}_\perp}
\newcommand{\jGW}{j^\mu_{\text {\tiny Goldstone-Wilczek}}}

\title{Axion~string~signatures~II:~A~cosmological~plasma~collider}

\date{\today}
\author[a]{Prateek Agrawal,}
\author[b]{Anson Hook,}
\author[c]{Junwu Huang,}
\author[b]{and Gustavo Marques-Tavares}

\affiliation[a]{Rudolf Peierls Centre for Theoretical Physics, Clarendon Laboratory, Parks Road, Oxford OX1 3PU, UK}
\affiliation[b]{Maryland Center for Fundamental Physics, University of Maryland, College Park, MD 20742, USA}
\affiliation[c]{Perimeter Institute for Theoretical Physics, 31 Caroline St.~N., Waterloo, Ontario N2L 2Y5, Canada}

\emailAdd{prateek.agrawal@physics.ox.ac.uk}
\emailAdd{hook@umd.edu}
\emailAdd{jhuang@perimeterinstitute.ca}
\emailAdd{gusmt@umd.edu}

\abstract{ We study early and late time signatures of both QCD axion
  strings and hyperlight axion strings (axiverse strings).  We focus
  on charge deposition onto axion strings from electromagnetic fields and
  subsequent novel neutralizing
mechanisms due to bound state formation.  While early universe
signatures appear unlikely, there are a plethora of late time
signatures.  Axion strings passing through galaxies obtain a huge
charge density, which is neutralized by a dense plasma of bound state
Standard Model particles forming a one dimensional ``atom''. The
charged wave packets on the string, as well as the dense plasma
outside, travel at nearly the speed of light along the string. These
packets of high energy plasma collide with a center of mass energy of
up to $10^{9}$ GeV. These collisions can have luminosities up to seven
orders of magnitude larger than the solar luminosity, and last for
thousands of years, making them visible at radio telescopes even when
they occur cosmologically far away.
The new observables are complementary to the CMB observables for
hyperlight axion strings that have been recently proposed, and are
sensitive to a similar motivated parameter range.}

\preprint{\today}

\begin{document}

\maketitle

\section{Introduction}\label{sec:intro}
Axions are one of the most compelling candidates for new physics~\cite{axion3,axion1,axion2,Svrcek:2006yi,Arvanitaki:2009fg}. They
are robustly predicted in many UV completions of the Standard Model,
and can naturally solve the strong CP
problem~\cite{axion3,axion1,axion2} and be dark matter.
In addition to solving these IR problems, they are also ideal
tools to access information about the UV. The axion coupling to
gauge bosons is topological in nature, and retains deep information
about the UV physics unpolluted by
intervening dynamics~\cite{Agrawal:2019lkr}. In this paper, we show that there is an additional way
that an axion can unlock UV physics -- by setting up an astrophysical
collider with center of mass energies as high as
$10^{9}$ GeV.

Axion strings are topological defects that provide an ideal avenue to
access UV information~\cite{Naculich:1987ci,Kaplan:1987kh,Manohar:1988gv,Harvey:1988in,Agrawal:2019lkr}. 
They can easily form in the early universe through the Kibble mechanism~\cite{Kibble:1976sj,Kibble:1980mv} as long as the temperature was high enough to restore the global symmetry
associated with the axion, typically called the Peccei-Quinn (PQ) symmetry.
The topological nature of the axion-photon
coupling leads to a number of model-independent effects.
In Ref.~\cite{Agrawal:2019lkr}, some of us explored how the presence of a
string affects the propagation of photons, and leaves a unique pattern in the cosmic microwave background (CMB). In this paper, we study how photons, or more precisely electromagnetic fields in the universe, affect axion string dynamics.

The two most motivated types of axion strings are the QCD axion strings and hyperlight axion strings. The QCD axion solves the strong CP problem~\cite{axion3,axion1,axion2} and its mass comes from QCD
dynamics and is essentially fixed in terms of its decay constant (see~\cite{diCortona:2015ldu} for more details). QCD axion strings can contribute to the abundance of axion dark matter~\cite{Gorghetto:2018myk,Gorghetto:2020qws,Buschmann:2019icd}. However, they
disappear shortly after the QCD phase transition, well before Big Bang Nucleosynthesis occurs~\cite{Gorghetto:2018myk,Buschmann:2019icd,Gorghetto:2020qws}. On the other hand,
hyperlight axions -- axions whose masses are comparable to or smaller than the Hubble constant today -- are a generic
consequence of the string axiverse~\cite{Svrcek:2006yi,Arvanitaki:2009fg,Demirtas:2018akl}. Hyperlight axion strings can easily persist
until today, where they can give us interesting signals both in the CMB~\cite{Agrawal:2019lkr,LiangDai} as well as the present day universe.

The interaction between the axion string and electromagnetic fields is
described by the beautiful phenomenon of anomaly inflow~\cite{Callan:1984sa,Bardeen:1984pm,Harvey:2000yg}.  The string
has a chiral zero mode trapped on the string core that carries electric charge~\cite{Jackiw:1981ee} and
only moves in one direction.  Because this mode is massless, the
string is superconducting and acts as a Quantum Hall system~\cite{Stone:2012ud}.
Much of the physics associated with superconducting axion strings was well
understood in seminal papers appearing in the
1980s~\cite{Witten:1984eb,Naculich:1987ci,Kaplan:1987kh,Manohar:1988gv}.
We will review in detail those results in the modern context with a specific emphasis on QCD axion strings and hyperlight axion strings.

Axion strings can obtain a large charge density in the background of present day or early universe magnetic fields~\cite{Naculich:1987ci,Kaplan:1987kh,Manohar:1988gv}.
When a magnetic flux $\Delta \Phi$ passes into a string loop, an electric charge $Q \propto \Delta \Phi$ moves onto the axion string core from the axion field profile surrounding the axion string. 
Magnetic field, as well as early universe scattering processes, can lead to charged axion string cores that
can have macroscopic charge and current densities, which results in extremely large electromagnetic fields around the string.

Axion strings charged under electromagnetism {\it neutralize}
quickly due to electromagnetic breakdown as the charge per unit length $\lambda_Q$ builds up. 
The high electric fields of a charged axion string can easily ionize surrounding gas and cause rapid production of SM particles around the string core due to, for example, stimulated Schwinger pair
production~\cite{Schwinger:1951nm,Erber:1966vv,Monin:2010qj} and the Blandford-Znajek process~\cite{10.1093/mnras/179.3.433}. This populates a dense cloud of charged particles with density as large as $\lambda_Q^3$. The charged
particles form bound states with the string, giving rise to a neutral ``atom'' extended
in one dimension. When
the charge density $\lambda_Q \gtrsim m_{PQ}$, the typical energy for a bound state
particle exceeds the mass of PQ fermions off the string $m_{PQ}$.  Collisions between these bound state particle and the zero mode fermions can scatter the zero mode fermions, and therefore charge overdensities, efficiently into a bulk fermions, {\it discharging} the string.
Thus, the axion string core never obtains a charge density larger than
$m_{PQ}$, which is required to stabilize axion string loops, aka vortons~\cite{DAVIS1988485,Brandenberger:1996zp,Fukuda:2020kym}.
Therefore, macroscopic axionic vortons which couple to gauge fields with light charged matter cannot be stabilized by charge or current.  Axionic vortons coupled to
QCD and/or electromagnetism, unfortunately, are of this type and are hence are unstable. 

Hyperlight axion strings (axiverse strings) can survive until today and would
occasionally pass through a galaxy. This sets up local charge
overdensities on the string with sizes of order the diameter of the galaxy and charge densities up to $\lambda_Q\sim 10^9 \,{\rm GeV}$ traveling at nearly the speed of light
down the string. The charged bound states around the axion string travels along with the string localized charge overdensities. These particles in the bound state can have energy up to $\sim {\rm min}(10^9 \,{\rm GeV}, f_a)$ and when they collide with other overdensities, act as extremely high energy colliders with extraordinary luminosity. A typical collision can have luminosities up to seven orders of magnitude larger than the solar luminosity, lasting for thousands of years. These collisions
occur often enough and are visible enough that despite happening $\sim$ Gpc away, they could still be looked for with radio telescopes and surveys like FAST, SKA~\cite{7303195} and CHIME~\cite{Amiri:2017qtx}. Observation of axiverse strings in the form of bright radio sources and in the CMB with edge detection techniques~\cite{LiangDai} would allows us to perform a non-trivial cross check and determine the origin of two spectacular signals. 

The paper is organized as follows. In section~\ref{sec:naculich}, we
provide a review of the Goldstone-Wilczek current and anomaly inflow. In section~\ref{sec:galacticB}, we discuss how Witten's superconducting strings and axion strings obtain charge and current when moving through magnetic
fields. In section~\ref{sec:boundstate}, we discuss the neutralization of the axion string due to particle production and the formation of bound states.  In
section~\ref{sec:kamehamehas}, we discuss the evolution of an axion string in galactic magnetic fields, the spreading of the
charge on the string, signals from charged plasma collisions, as well as their observational prospects. Section~\ref{sec:conclusion} serves as a
conclusion. In appendix~\ref{sec:earlyuniverseB}, we discuss how the total charge of string - axion system can change and the corresponding observational prospects. In appendix~\ref{sec:QCD}, we briefly describe the evolution of QCD axion string loops in the early universe. In appendix~\ref{sec: littleparks}, we review some of the effects we discuss from a condensed matter perspective, and point out the key differences between our systems and the corresponding condensed matter systems.

\section{The Consistent and Covariant Anomaly on Axion Strings}
\label{sec:naculich}
In this section, we review the physics of an extended string.
There are many possible kinds of cosmic strings such as axion strings,
Witten's superconducting strings (Witten string) and local strings (Abrikosov-Nielsen-Olesen vortices).
Unless otherwise mentioned, the axion strings we will review have unit winding number and have
an electromagnetic anomaly coefficient of $\mathcal{A} = 1$ saturated by a single PQ fermion of electric charge 1. In this section, we focus on the case of $U(1)_{\rm EM}$. Extending this discussion to $U(1)$ hypercharge and the full SM gauge group in the UV is straight forward.
The anomaly coefficient is defined by the axion photon coupling
\bea
\mathcal{L} 
\supset 
- \frac{\anomaly e^2}{16 \pi^2 f_a} 
a F_{\mu \nu} \widetilde  F^{\mu \nu}
\eea
where $\widetilde F^{\mu \nu} = \frac12 \epsilon^{\mu \nu \rho \sigma} F_{\rho \sigma}$.

The light degrees of freedom on the string are almost completely fixed by
the phenomenon of anomaly inflow.
In particular, the anomalous axion coupling in the bulk dictates that a
charged chiral state must reside on the string. 
The electromagnetic current carried by the string is known to 
have a few subtleties~\cite{Naculich:1987ci,Kaplan:1987kh,Manohar:1988gv,Harvey:1988in,Stone:2012ud}.  In this section, we review the
physics of anomaly inflow and the difference between the consistent and
covariant anomalies that arise on the string.

To simplify matters, we will consider the theory of axion strings without
domain walls, namely we will take the axion to be effectively massless.
An infinite single string is topologically
stable and constitutes a saddle point in the path integral. Even
though a network
of strings or closed strings are not topologically stable, the
partition function will factorize into multi-string states in a dilute
string limit\footnote{In fact, a single isolated string has a
  log-divergent tension, and hence is not a finite action configuration.
  However, the existence of a string network or a finite
sized loop cuts off this logarithmic divergence.}.
To calculate the contribution of a single string, we expand the
action around the classical field configuration. We get a 1+1
dimensional field theory on the string, which is coupled to a 3+1
dimensional bulk field theory. We integrate out all massive degrees of
freedom, so that we are only left with the axion and the photon in the bulk
and massless modes required by index theorems on the axion string.

The action for the 3+1 dimensional anomalous field theory is,
\begin{align}
  S_{3+1}
  &=
  \frac{e^2}{16\pi^2 f_a} 
  \int  d^4 x (1+\rho)
  \epsilon^{\mu\nu\rho\sigma} \partial_\mu a A_\nu F_{\rho\sigma}
  \, ,
  \label{eq:axion-anomaly}
\end{align}
where $e$ is the electric charge, $a$ is the axion field, $A$ is the photon field, and $F$ is the electromagnetic field strength.
This form is similar to the more familiar form $a F \widetilde{F}$ up to
the presence of the ``bump function'' $\rho(r)$~\cite{Harvey:2000yg} and
integration by parts. $\rho(r)$ is present because string boundary
terms play an important role and $\rho(r)$ allows one to cleanly
separate out the fermion zero modes and the bulk wavefunctions in the
presence of the string. 
The function $\rho(r)$ 
parametrizes the embedding of the string in the 4D space and is related to
the wavefunction of the zero-mode fermion on the string, $F(r)$,
\begin{align}
  \rho (r) = -1+ 2 \pi \int_0^r d\sigma \sigma F^2(\sigma)
  \label{eq:rhor}
\end{align}
From this definition and the normalization condition for the fermion
zero mode $\int \sigma d\sigma d\phi F^2(\sigma) = 1$, we see that
\begin{align}
  \rho(0) &= -1, \qquad
  \rho(r\to \infty) = 0,
  \\
  \rho'(0) &= \rho'(\infty)=0,
  \label{eq:rho}
\end{align}
where $r=0$ is the location of the center of the string core. The bump function and the axion field profile around the
string core provide a smoothed $\delta$-function centered at the
location of the string core,
\begin{align}
  \frac{1}{2\pi f_a}
  \partial_r \rho\, \partial_\phi a
  \equiv
  \delta^{(2)}(\xperp)
  \,,
  \label{eq:bump-two-form}
\end{align}
where $\xperp$ is a 2-vector normal to the string worldsheet.

In axion electrodynamics, it is well-known that gradients of axions in
the presence of magnetic fields carry electromagnetic
charge~\cite{Goldstone:1981kk,Wilczek:1987mv}.
This current can be found from the action $S_{3+1}$,
\begin{align}
  j^\mu_{\text {3+1}}
  &=
  \jGW
  + j^\mu_{\text {\tiny A}}
  \\
  \jGW
  &=
  \frac{e^2}{8\pi^2 f_a} (1+\rho) 
  \epsilon^{\mu\nu\alpha\beta} 
  \partial_\nu a F_{\alpha \beta} 
  \\
  j^\mu_{\text {\tiny A}}
  &=
  \frac{e^2}{8\pi^2 f_a} 
  \epsilon^{\mu\nu\alpha\beta} 
  \partial_\alpha \rho\, \partial_\nu a A_\beta
  \approx
  -\frac{e^2}{4\pi} 
  \delta^{2}(\xperp)
  \epsilon^{\mu\nu} 
  A_\nu ,
  \label{eq:GWcurrent}
\end{align}
where we have defined 
$\epsilon^{\mu \nu} = \epsilon^{a b}$ for $\mu, \nu = {t,z}$ and 0 otherwise. We are using the notation where the Greek indices run over the entire
3+1 dimensional spacetime while the indices $a,b$ run over the 1+1
dimensional spacetime of the string core.
The first equation is the total current derived from the bulk action.
The second equation defines the Goldstone-Wilczek
current~\cite{Goldstone:1981kk} or
the Hall current, which has a familiar form that describes a current
off of the string core. 
The last equation denotes a string localized current carried by the axion field
and its radial mode.
To track conservation of electric charge, we calculate the divergence 
of each of the contributions to the total current, 
\begin{align}
  \partial_\mu \jGW
  &=
  \frac{e^2}{8\pi^2 f_a} 
  \epsilon^{\mu\nu\alpha\beta} 
  \partial_\mu  \rho \partial_\nu a F_{\alpha \beta}
  =
  \delta^{(2)}(\xperp)
  \frac{e^2}{4\pi} \epsilon_{ab}F^{ab}
  \\
  \partial_\mu j^\mu_{\text A}
  &=
  -
  \delta^{(2)}(\xperp)
  \frac{e^2}{8\pi} \epsilon_{ab}F^{ab}
  \,.
  \label{eq:GWdivstring}
\end{align}
Note the famous factor of
two difference between the final result for $\partial_\mu j^\mu_{3+1}$ and 
the divergence of the Goldstone-Wilczek current.

The 1+1 theory on the string core itself has a electromagnetically charged chiral fermion zero mode.
Chiral fermion zero modes have an anomaly, and hence violate charge
and energy conservation. The 1+1 dimensional action is
\bea
S_{1+1} = \int d^2 x \, \left ( \,  i  \bar{\psi} ( \slashed{\partial} - i e \slashed{A} ) \psi - \frac{e^2}{4 \pi} A_a A^a\, \right ) .
\eea
The second term is a counter term that is required to match anomalies.
Some more physical intuition for this counter term can be obtained by remembering
that anomalies give gauge bosons a mass.  Since the photon is massless, a negative mass squared for the photon is required on the string to cancel the anomaly's contribution to the photon mass. 
The electric currents carried by the fermionic zero mode and the counter term are
\begin{align}
  j^a_{\rm zero-mode}
  &=
  e \bar{\psi}\gamma^a \psi \qquad j^a_{\rm c.t.}  = - \frac{e^2}{4 \pi} A^a.
  \label{eq:zero-mode-current}
\end{align}
These currents are anomalous and obey
\begin{align}
  \partial_a j^a_{\rm zero-mode}
  &=
  -\frac{e^2}{4\pi} \left ( \partial_t + \partial_z \right ) \left ( A_t - A_z \right ) \qquad \partial_a j^a_{\rm c.t.}
  =
  -\frac{e^2}{4\pi} \partial_a A^a  \\
  \partial_a j^a_{\rm 1+1}
  &= \partial_a j^a_{\rm zero-mode} + \partial_a j^a_{\rm c.t.} = 
  -\frac{e^2}{8\pi} \epsilon^{ab}F_{ab} .
  \label{eq:consistent-anomaly}
\end{align}
This is sometimes referred to as the consistent
anomaly since it satisfies the Wess-Zumino consistency condition on
the gauge variation of the effective action $W$,
  $(\delta_\epsilon \delta_\eta - \delta_\eta
\delta_\epsilon) W = \delta_{[\epsilon,\eta]} W$. This automatically
follows from the definition of this anomaly as a local functional
variation of the action.

As we saw above, the divergence of the Goldstone-Wilczek current by itself 
is twice the consistent anomaly, and the extra string localized contribution to the
current $j^\mu_{\text{A}}$ is crucial to ensure that the
total anomaly cancels and electromagnetic current is conserved.  This extra contribution to the current is peaked at
the string core.  Therefore, the current associated with the string can be
thought of as the zero mode current plus this extra contribution from
the boundary variation of the bulk effective action. 
\begin{align}
  j^\mu_{\rm string}
  &=
  j^\mu_{\rm zero-mode} + j^\mu_{\rm c.t.}  + j^\mu_{A}.
  \label{eq:string-current}
\end{align}
The divergence of this combination is 
\bea \label{Eq: covariant}
  \partial_\mu j^\mu_{\rm string} = -\frac{e^2}{4\pi} \epsilon^{ab}F_{ab}
\eea
and is called the covariant anomaly.  The name covariant anomaly stems from
the non-abelian version of the anomaly where the consistent anomaly
again involves $\partial_\mu j^{a,\mu}_{\rm string}$ while the name ``covariant anomaly'' comes from the fact that the non-abelian version of Eq.~\ref{Eq: covariant} involves $D_\mu j^{a,\mu}_{\rm string}$ and is gauge covariant.  In the context of an abelian anomaly, the only difference is a factor of 2 and the definition of the current.

The divergence of the total current on the string
cancels against the
anomaly in the Goldstone-Wilczek current.
\begin{align}
  \partial_\mu j^\mu_{\rm total} = \partial_\mu \jGW  + \partial_\mu j^\mu_{\rm string}
  &=
 \partial_\mu \jGW  + \partial _\mu j^\mu_{\rm zero-mode} + \partial _\mu j^\mu_{\rm c.t.} + \partial_\mu j^\mu_{A}
  = 0.
  \label{eq:consistent-covariant}
\end{align}
This is just another equivalent way of seeing that the total anomaly
cancels and that the electromagnetic current is conserved.

The electromagnetic current on the string is carried
by the fermionic zero mode as well as a combination of the axion and the gauge field.
The zero-mode on the string is massless and moves at the speed of
light, therefore, the charge density and the electric current for that mode
are correlated,
\begin{align}
  \lambda_{Q,\,\text{\tiny zero-mode}}
  &=
  I_{\text{\tiny zero-mode}} .
  \label{eq:zero-mode-charge-current}
\end{align}
There is another charge carrier in the form of the currents $j^\mu_a = j^\mu_{\rm c.t.} + j^\mu_{A}$.  This combination also has
\begin{align}
  \lambda_{Q,\,\text{\tiny a}}
  &=
  I_{\text{\tiny a}} .
\end{align}
One typically lumps all of these contributions together and simply states that there is a single fermion that lives on the string that moves at the speed of light.  The logic for grouping all of the three currents together is that none of the three currents are separately gauge invariant and only the combination is physical.  Thus in this simple example we have 
$\lambda_{Q,\,\text{\tiny string}} = I_{\text{\tiny string}}$.

The equality of charge density and current is not true in general and is due to that the anomalies were matched with only a single left moving fermion.  If the anomaly matched with a combination of left movers and right movers, then the situation would be different.  Right movers obey $\lambda_{Q,\,\text{\tiny right}} = - I_{\text{\tiny right}}$ so that one would have $\lambda_{Q,\,\text{\tiny string}} \ne I_{\text{\tiny string}}$.

Our discussion is simple to extend to the case
where the axion is massive. The axion profile around
the string core forms a domain wall (or multiple domain walls) ending on
the string core. We can proceed to integrate out the axion as well, and
study the effective 2+1 dimensional theory. 
This domain-wall + string system is the familiar Quantum Hall
system, with a Chern-Simons theory living on the domain wall and
chiral edge
states living on the string core.  The physics of anomaly inflow
in this case is exactly as described above~\cite{Stone:2012ud}.

\section{String - Magnetic field interactions} 
\label{sec:galacticB}

In this section, we describe how strings obtain charge and current
through their interaction with magnetic fields.  We first review how
Witten's superconducting strings gain and lose current before proceeding to the case 
of the axion string. In the process, we highlight a key difference between
the Witten string, which is essentially magnetic, and the axion string, which is
essentially electric.  This difference can lead to drastically different phenomenology.

In order to study the effects of magnetic fields, it will be crucial that
magnetic fields can pass into string loops.  Namely, flux through a loop is not
conserved for both Witten strings and axion strings. We refer to this phenomenon as {\it flux non-conservation}.
For most superconductors found in a lab, flux through a
superconducting loop is quantized and in most circumstances conserved.
Magnetic flux can enter or leave a superconducting ring only by
tunneling vortices that carry unit magnetic flux through the
superconducting bulk.  As will be proven below, 
Witten and axion strings do not have a conserved flux because 
they are so narrow that magnetic flux can jump
across the string easily (see Appendix~\ref{sec: littleparks} for a
detailed discussion of the non conservation and non quantization of
the magnetic flux from a more condensed matter perspective).
Flux non-conservation will be critical for axion strings as flux
non-conservation is related to charge non-conservation so that
axion strings also acquire electric charge as magnetic fields
pass by.

\subsection{Charge conservation}
We will be specifically interested in the charge/current on a string core and
how it changes in time and how it moves in space.  The most useful way
to track the total charge of
a string is in terms of flux.  The reason for this is Faraday's Law.
The change in the charge of a string can be re-expressed using the
anomaly to be 
\bea
\Delta Q &=& \int dt \dot Q = \int d^2x \frac{d \lambda_Q}{dt} = \int d^2x \partial_a j^a_{\rm string} = -\frac{e^2}{4\pi} \int d^2 x \epsilon^{ab}F_{ab} \nonumber \\
&=& -\frac{e^2}{2 \pi} \int d^2 x E =  \frac{e^2}{2 \pi} \int dt \frac{d \Phi}{dt} = \frac{e^2 \Delta \Phi}{2 \pi}, \label{Eq: delta Q}
\eea
where we have assumed that the magnetic field is uniform along the string direction. If one neglects particles being able to go onto or off of the string core through means other than electric fields, then one obtains
\bea
\label{Eq: holy grail}
Q_{\text{string}} =  \frac{e^2 \Phi}{2 \pi} .
\eea

In a more general setting (see appendix~\ref{sec:earlyuniverseB} for more details), we have
\bea
\label{Eq: sum}
Q_{\text {axion}} + Q_{\text{string}} =  Q_\text{tot},
\eea
where $Q_\text{tot}$ is the total charge that is carried by the axion string core together with the fluffy axion cloud around it. The charge stored in the axion field can be easily found to be
\bea
Q_{\rm axion} = Q_{\text {\tiny Goldstone-Wilczek}} = - \frac{e^2 \Phi}{2 \pi} .
\eea
We thus find the simple formula
\bea
Q_{\text{string}} =  Q_\text{tot} + \frac{e^2 \Phi}{2 \pi} ,
\eea
generalizing Eq.~\ref{Eq: holy grail}. This relation is the same as the one derived in~\cite{Naculich:1987ci}. The charge $Q_\text{tot}$ is quantized and conserved in most circumstances relevant for our future discussions. We leave the various cases where $Q_\text{tot}$ can change to appendix~\ref{sec:earlyuniverseB}.

\subsection{Witten's superconducting strings moving through magnetic fields} 
\label{Sec: witten mag}

We will be interested in how axion strings acquire charge and currents as an
they move through a magnetic field.  
To demonstrate the power
of Eq.~\ref{Eq: holy grail}, we first reproduce how current goes onto
Witten strings when they are passing through magnetic fields before
showing how the result was originally derived.
A Witten string can be thought of as two different axion strings of
opposite orientation placed
on top of each other so that the anomalies cancel.  We can use intuition from the axion
string example as long as one restricts to $j^\mu_{\rm zero-mode}$ and forgets
about $j^\mu_A$ as the $j^\mu_A$ from the two axion strings cancel
each other.  

\subsubsection{How Witten's superconducting strings obtain current}

As first shown in Ref.~\cite{Witten:1984eb}, when Witten strings pass
through a magnetic field $B_0$, particles are produced on the wire at
a rate
\bea
\label{Eq: witten}
\frac{dN}{dt dz} = \frac{e \eta}{2 \pi} v B_0 .
\eea
We derive this equation in two different ways below -- using 
Eq.~\ref{Eq: holy grail} as well as considering the scattering of
photons off of the string. From Eq.~\ref{Eq: holy grail} it follows that
\bea
\label{Eq: dN}
\frac{dN}{dt dz} = e \frac{d \Phi}{dt dz} .
\eea
Thus the problem boils down to understanding what is the total flux that passes through the loop when there is an applied magnetic field
$B_0$.  The
reason why this is non-trivial is because of self inductance.  
As shown before, the charge per unit length on the loop is
\bea
\lambda_Q^{(1)} =  \frac{e^2 }{2 \pi}\frac{d \Phi}{d z} \qquad 
\lambda_Q^{(2)} =  -\frac{e^2 }{2 \pi}\frac{d \Phi}{dz} ,
\eea
where the superscripts indicate the two opposite charged zero modes
living on the string.  A Witten string has two zero modes as opposed
to the axion's single zero mode and the
anomalies of the two zero modes cancel each other, that is $\lambda_Q^{(1)}
+ \lambda_Q^{(2)} =0$.
The total current on the string is
\bea
I = I_1 + I_2 = \lambda_Q^{(1)} - \lambda_Q^{(2)} = \frac{e^2 }{\pi}\frac{d
\Phi}{dz} ,
\eea
where the minus sign comes from the fact that the two zero modes
travel in opposite directions.
The magnetic flux from the current $I$ is 
\bea
\frac{d \Phi_I}{dz} = \frac{I}{2 \pi} \log (\Lambda/\omega)
\eea
where we have introduced the size of the string core $\sim 1/\Lambda$ and the oscillation
frequency of the applied B field $\omega$ to regulate the log divergent magnetic flux similar to~\cite{Witten:1984eb}.  The
total flux that goes through the string is
\bea
\label{Eq: flux}
\Phi = \Phi_0 - \Phi_I \qquad \rightarrow \qquad \Phi = \frac{\Phi_0}{1 + \frac{e^2}{2 \pi^2} \log (\Lambda/\omega)} , 
\eea
where the minus sign comes from Lenz's law.  Using Eq.~\ref{Eq: dN} and Eq.~\ref{Eq: flux} we
reproduce Witten's result using the anomaly equation. Moreover, the above mentioned computation allows us to calculate $\Lambda$ explicitly in the case of a circular loop with a uniform current. 
In this situation,
\begin{equation}
\frac{\Lambda}{\omega} = \frac{8}{e^2} \frac{R}{r_s}
\end{equation}
where the $e$ in this equation is Euler's number rather than electric charge, $r_s$ is the radius of the string and R is the radius of the string loop.

As it will be relevant for axion strings, we will review how Eq.~\ref{Eq: witten} was originally derived.
The bosonized action for the zero-mode fermions is~\cite{Witten:1984eb}
\bea
S = \int dz \,dt \,  \left [ \frac{1}{2} (\partial_i \phi)^2 - \frac{e}{\sqrt{\pi}} \phi E \right ]
\eea
where E is the electric field on the string.  Consider scattering a photon heading straight towards the string and polarized in the z direction.  We choose a gauge where $A_t = A_x = A_y = 0$, $A_z = A(x,y,t)$ and $\phi = \phi(t)$. 
The equations of motion are
\bea
\ddot A - \nabla^2 A - \frac{e}{\sqrt{\pi}} \delta^2(x) \dot \phi = 0 \\
\ddot \phi + \frac{e}{\sqrt{\pi}} \dot A(x=y=0) = 0
\eea
These can be simplified to
\bea
\label{Eq: witten scatter}
\omega^2 A  = - \nabla^2 A + \frac{e^2}{\pi} \delta^2(x) A . 
\eea
This 2 dimensional scattering problem can be solved exactly and one finds
that at the origin
\bea
\frac{A(0)}{A_\text{incident}} 
= \eta = \frac{1}{1 + \frac{e^2}{2 \pi^2} \log (\Lambda/\omega) } .
\eea
This shows that the electric field on the string is screened to be a smaller value by an amount $\eta$.
Using this, we find that
\bea
\frac{dN}{dt dz} 
= e \frac{d \Phi}{dt dz}  = \frac{e E}{2 \pi} = \frac{e \eta}{2 \pi} v B_0 ,
\eea
proving Eq.~\ref{Eq: witten}.
Note that the presence of the large magnetic field will have
phenomenological effects
analogous to the magnetic black hole studied in~\cite{Maldacena:2020skw}, but
extended along one dimension.

\subsubsection{Flux non-conservation}
\label{Sec: witten nonflux}

Flux non-conservation was made precise for Witten's superconducting strings
in~\cite{Witten:1984eb}.  We can continue solving the scattering process started in 
Eq.~\ref{Eq: witten scatter} to find that photons pass through
the superconducting string with only a small differential scattering cross section
of
\bea
\frac{d \sigma}{d z} = \frac{e^4 \eta^2}{8 \pi^3} \frac{2 \pi}{\omega},
\eea
where $z$ is the coordinate along the string direction, $\omega$ is the photon angular frequency and $e$ is the electric charge of the fermion zero-mode living on the string.
The fact
that this differential cross section is much smaller than the wavelength of the photon $\lambda= 2\pi/\omega$, means that most photons, and
in particular magnetic flux, will pass through the loop without noticing the existence of the string, and thus we have flux non-conservation.

\subsection{Axion strings moving through magnetic fields}

In this subsection, we move on to describe how axion strings gain and
lose charge.
Axion strings can be studied in much the same way as
Section~\ref{Sec: witten mag}.
An extremely important difference is that while
Witten strings gain current but no charge, axion strings gain both charge
and current as they cross magnetic fields.
This results in a
sizeable electric field induced by the string which affects the
phenomenology significantly.

We can see the relation between charge and current in an explicit
example.  As a simplification, take the axion to have a mass so that
there is a domain wall filling in the axion string loop.  This
assumption is not necessary but simplifies calculations.  We will
consider the case of a domain wall and string system that has a total
charge
\bea
Q_{\text{DW}} + Q_{\text{string}} =  Q_\text{tot} .
\eea
The charge on the domain wall coming from the Goldstone-Wilczek
current can be found to be
\bea
Q_{\text{DW}} = - \frac{e^2 \Phi}{2 \pi} = - \frac{e^2}{2 \pi}  R I_{\text{string}} \log (\Lambda R) ,
\eea
where $\Phi$ is the flux inside the string loop, $R$ is the radius of the string loop 
and $I_{\text{string}}$ is its current.
Using conservation of charge, we find that charge and current are related by 
\bea
Q_{\text{string}} = Q_\text{tot} + \frac{e^2 R I_{\text{string}} }{2 \pi} \log (\Lambda R) .
\eea
In the simple example where charge and current are related, this allows one to solve for the charge on the string in terms of the total charge.  However, in a more general situation, the two are not so easily related to each other.

\subsubsection{How axion strings obtain charge and current}

We now consider the more complicated case of what happens to an axion string
when flux crosses it.
We will follow Witten's calculation and use scattering to find the charge and current on the string.
In the case of axions, the bosonized action for the zero-mode fermion is~\cite{Naculich:1987ci}
\bea
S = \int dz \,dt \, \left [ \frac{1}{2} (\partial_i \phi - \frac{e}{2 \sqrt{\pi}} A_i)^2 - \frac{e}{2 \sqrt{\pi}} \phi E \right ]
\eea
where E is the electric field on the string.  
As before, consider scattering a photon heading straight towards the string and polarized 
in the z direction.  We will again choose a gauge where $A_t = A_x = A_y = 0$, 
$A_z = A(x,y,t)$ and $\phi = \phi(t)$. 
The equations of motion are
\bea
\ddot A - \nabla^2 A - \frac{e}{2 \sqrt{\pi}} \delta^2(x) \dot \phi + \frac{e^2}{4 \pi} \delta^2(x) A = 0 \\
\ddot \phi + \frac{e}{2 \sqrt{\pi}} \dot A(x=y=0) = 0
\eea
Note that we are neglecting the effect of the Goldstone-Wilczek current.  The reason for this is that
as is well known, the effect of this current is to rotate the
polarization of the incident electric and magnetic
fields.  As all it does is change how the photons propagate to and away from the string, we will ignore it
in much the same way that scattering cross sections consider only amputated diagrams (see Ref.~\cite{Agrawal:2019lkr} for the observable effects of this polarization rotation).
The equations of motion can be simplified to
\bea
\omega^2 A  = - \nabla^2 A + \frac{e^2}{2 \pi} \delta^2(x) A . 
\eea
This two-dimensional scattering problem can be solved exactly and one finds
that at the origin
\bea
\label{Eq: screen}
\frac{A(0)}{A_\text{incident}} = \eta' = \frac{1}{1 + \frac{e^2}{4 \pi^2} \log (\Lambda/\omega) } .
\eea
This shows that the electric field on the string is screened to be a smaller value by an amount $\eta'$.

As with the case of the Witten string, we can derive this result using fluxes, currents and
charges.  Let us consider an axion string with zero initial charge or current which passes through 
a magnetic field region with flux $\Phi_0$, the charge per unit length on the loop is (following~\ref{Eq: dN}):
\bea
\lambda_Q^{(\rm string)} =  \frac{e^2 }{2 \pi}\frac{d \Phi}{d z}.
\eea
As before, the magnetic flux from the current $I$ is 
\bea
\label{Eq: dos}
\frac{d \Phi_I}{dz} = \frac{I}{2 \pi} \log (\Lambda/\omega) .
\eea
Following equations~\ref{Eq: flux} and~\ref{Eq: screen}, we have
\bea
\label{Eq: tres}
\frac{d \Phi}{d z} = \frac{d \Phi_0}{dz} - \frac{d \Phi_I}{dz} \quad \mathrm{and} \quad \frac{d \Phi}{d z}  = \eta' \frac{d \Phi_0}{d z} ,
\eea
where the first equality is the definition of total flux and the second equality comes from our scattering result.  
%

To summarize, we show that if an axion string passes through
a region of space with a magnetic field $B$ and physical size $d$ (up to geometric $\mathcal{O}(1)$ numbers)
\bea
\frac{dQ_\text{string}}{dt dz} = e^2 \frac{d \Phi}{dt dz}  = \frac{e^2 \eta'}{2 \pi} v B  \\
Q_\text{string} = \frac{e^2 \eta'}{2 \pi} B d^2 \qquad \lambda_Q^\text{string} = \frac{e^2 \eta'}{2 \pi} B d v_s .
\eea
For the systems we consider in the rest of this paper, this parameter
\begin{equation}
\eta' \approx \frac{1}{1+ \frac{\mathcal{A} e^2}{4 \pi^2} \log (f_a/H_0) } \approx 0.8
\end{equation}
for an anomaly coefficient $\mathcal{A} \simeq 1$. In the following sections, we will approximate $\eta' = 1$ unless otherwise mentioned.

\subsubsection{Flux/Charge non-conservation}

As before, we can continue the scattering calculation to find that
the cross section of a photon with an axion string is 
\bea
\label{Eq: photon axion scatter}
\frac{d \sigma}{d z} = \frac{e^4 \eta'^2}{8 \pi^3} \frac{2 \pi}{\omega},
\eea
where $\omega$ is the photon angular frequency and $e$ is the electric charge of the fermion zero-mode living on the string.
Again, we find that the differential cross section is much smaller
than the wavelength of the photon $\lambda= 2\pi/\omega$, so that
magnetic flux can also pass through the axion string loop. This flux
non-conservation for axion string loops also reinforces the fact that
the electric charge in the fluffy axion cloud around the axion string
core, as well as the axion string core itself, is not individually
conserved or quantized.

\section{Charged strings and bound states}\label{sec:boundstate}

As will be discussed more explicitly in section~\ref{sec:kamehamehas}, axion strings can carry electric charge densities and currents as large as $\lambda_Q \sim 10^{9}$ GeV coming from interactions with present day magnetic fields and possibly even larger from interactions with early universe magnetic fields.
These electrically and magnetically charged axion strings or charged sections of axion strings can carry a huge electric field
\begin{equation}
\vec{E} = \frac{\lambda_Q}{2 \pi r} \hat{r},
\end{equation}
which is large enough to ionize all surrounding Standard Model matter
and lead to many instabilities, such as stimulated Schwinger pair production of all of the known charged
particles in a very wide range of distances from the string core. 
In this section, we discuss how particle production {\it neutralizes} the enormous charge densities on the axion string core due to bound state formation.

\subsection{Neutralization of an axion string due to bound states} 
\label{Sec: gmt}

As a warm-up for the more complicated case of an axion string with both electric and magnetic fields, we first consider the case of a charged wire with no current.
Consider a system where the charge
per unit length at a distance $r$ from the string is given by
$\lambda_Q^{\rm eff}(r)$ (similar to the effective charge
of the nucleus in atomic physics). 
The bound states of this 2 + 1 dimensional system satisfy 
\bea
\label{Eq: bound states old}
\lambda_Q^{\rm eff} (r) \gg m_{\rm SM} &\qquad&  \langle r \rangle \sim \frac{n}{\lambda_Q^{\rm eff} (r)} \qquad  \qquad \, \, \, \, \,\, E_\text{bound} \sim \frac{e \lambda_Q^{\rm eff} (r)}{2\pi} \log \left ( \frac{\langle r \rangle}{R} \right ) \\
\lambda_Q^{\rm eff} (r) \ll m_{\rm SM} &\qquad&  \langle r \rangle \sim \frac{n}{\sqrt{\lambda_Q^{\rm eff} (r) m_{\rm SM}}} \qquad E_\text{bound} \sim \frac{e \lambda_Q^{\rm eff} (r)}{2\pi} \log \left ( \frac{\langle r \rangle}{R} \right ), \nonumber
\eea
where $m_{\rm SM}$ is the mass of the Standard Model
particle bound to the string core, $R$ is the length of the axion string, and $n$ is an integer.  As long as $\lambda_Q^{\rm eff}
(r) \gg m_{\rm SM}$,  the bound state energy is negative enough to
make it 
preferable to pair produce bound state particles~\footnote{More
precisely the pair production produces a bound state particle with a
free antiparticle or visa versa.}.  This pair production occurs at a
rate of $dN/dAdt \sim \lambda_Q^3$, where $A$ is the cross section area, if a naked charged string core ever
exists and quickly fills up the 
bound states of the 2 + 1 dimensional ``atom".

We are now ready to move on to the system at hand, an axion string whose current and charge density are equal. Note that classicaly the $E$ and $B$ fields would both work to confine the motion of particles with charge opposite to the string to be close to the string. In fact, a particle at rest would fall into the string and get accelerated parallel to the string asymptotically approaching the core of the string. To find the quantum bound states we can consider wave-functions of the form $\exp(i k_z z) \phi(r, \theta)$, where we took the direction of the string to be along the $z$ axis, and use the WKB approximation for the bound states of the 2 dimensional system showing that they satisfy 
\bea
\label{Eq: bound states}
\lambda_Q^{\rm eff} (r) \gg m_{\rm SM} &\qquad&  \langle r \rangle \sim \frac{n}{\lambda_Q^{\rm eff} (r)} \sqrt{\frac{\lambda_Q^{\rm eff} (r)}{E_\text{bound} - k_z}} \qquad   \, \, \, \,  E_\text{bound} \sim \frac{e \lambda_Q^{\rm eff} (r)}{2\pi} \log \left ( \frac{\langle r \rangle}{R}   \right ) - k_z  \\
\lambda_Q^{\rm eff} (r) \ll m_{\rm SM} &\qquad&  \langle r \rangle \sim \frac{n}{\sqrt{\lambda_Q^{\rm eff} (r) m_{\rm SM}}} \qquad \qquad \qquad E_\text{bound} \sim \frac{e \lambda_Q^{\rm eff} (r)}{2\pi} \log \left ( \frac{\langle r \rangle}{R} \right ), \nonumber
\eea
There are several things to note about these estimates. First, note that despite the suggestive form, $k_z$ does not correspond to physical momentum due to the presence of the magnetic field and the bound states are not momentum eigenstates. Second,  unless $E_\text{bound} \approx k_z$, the bound state properties are essentially independent of the magnetic field.  This property can be easily seen by comparing the electric and magnetic forces with each other.  Another thing to note is that the bound states with positive $k_z$ have smaller bound state energies.  This leads to the expected result that the generic bound state will have significant momentum in the direction of the current, and thus screen both the electric and magnetic fields.

In such a large electric field, standard model particles are
accelerated to energies logarithmically larger than $\lambda_Q$ over a
very short distance, potentially leading to a bright cosmic ray
source. However, similarly to the Blandford-Znajek effect, these
accelerated fermions will instead lead to a runaway process of
stimulated pair production where a huge density of high energy
SM particles are created.  The particles that carry a
charge that is the same sign as the charge of the string core will be
ejected while the particles that carry the opposite sign charge as the
string core will be captured by the string core into tightly bounded
orbits with properties shown in Eq.~\ref{Eq: bound states}.  The axion
string and plasma bound states behaves like a large one-dimensional
atom which is only very mildly ionized, similarly to the ionized gas in
the universe.

As a result of this extremely quick neutralization process, unless the
string accidentally came across a stellar object with a large magnetic
field, it would always stay in an approximately steady state as it proceeds
through the galactic magnetic field. These B-fields charge up the
string at a rate
\begin{equation}
\frac{\mathrm{d} \lambda_Q}{\mathrm{d}t} \approx \alpha B v_s^2 .
\end{equation}
As the string core acquires larger charge, the combined bound state
system retains the same charge while charged particles with the same
charge as the string are emitted from the system at the same rate.

In the process of charging up the string (or discharging the string when the string encounters a magnetic field region with the opposite orientation), the charge density on the string core changes.  Such an effect is similar to the $\beta$ or inverse $\beta$ decay of an atom, which will lead to subsequent $\gamma$ decay as the ``electron cloud" surrounding the string core rearranges itself. In this case, given the huge binding energy of the charged particles, the emitted radiation can be in the form of neutrinos, pions and even electroweak gauge bosons. However, these high energy emissions will be mainly reabsorbed by the extremely dense cloud of charged particles around the string core and thermalize, since the rate at which the string core acquires charge ${\rm d}\lambda_Q/ {\rm d}t \sim \alpha B v_s$ is quite small. 

We expect, as a result, that the axion string core is coated by a dense plasma of extremely high energy standard model particles which is emitting both charged and neutral radiation mainly from its surface. In order to understand this surface emission, we will need to understand the structure of the plasma outside of the string core. 

\subsection{The structure of a charged axion string}

\begin{figure}
\centering
\includegraphics[width=0.8\textwidth]{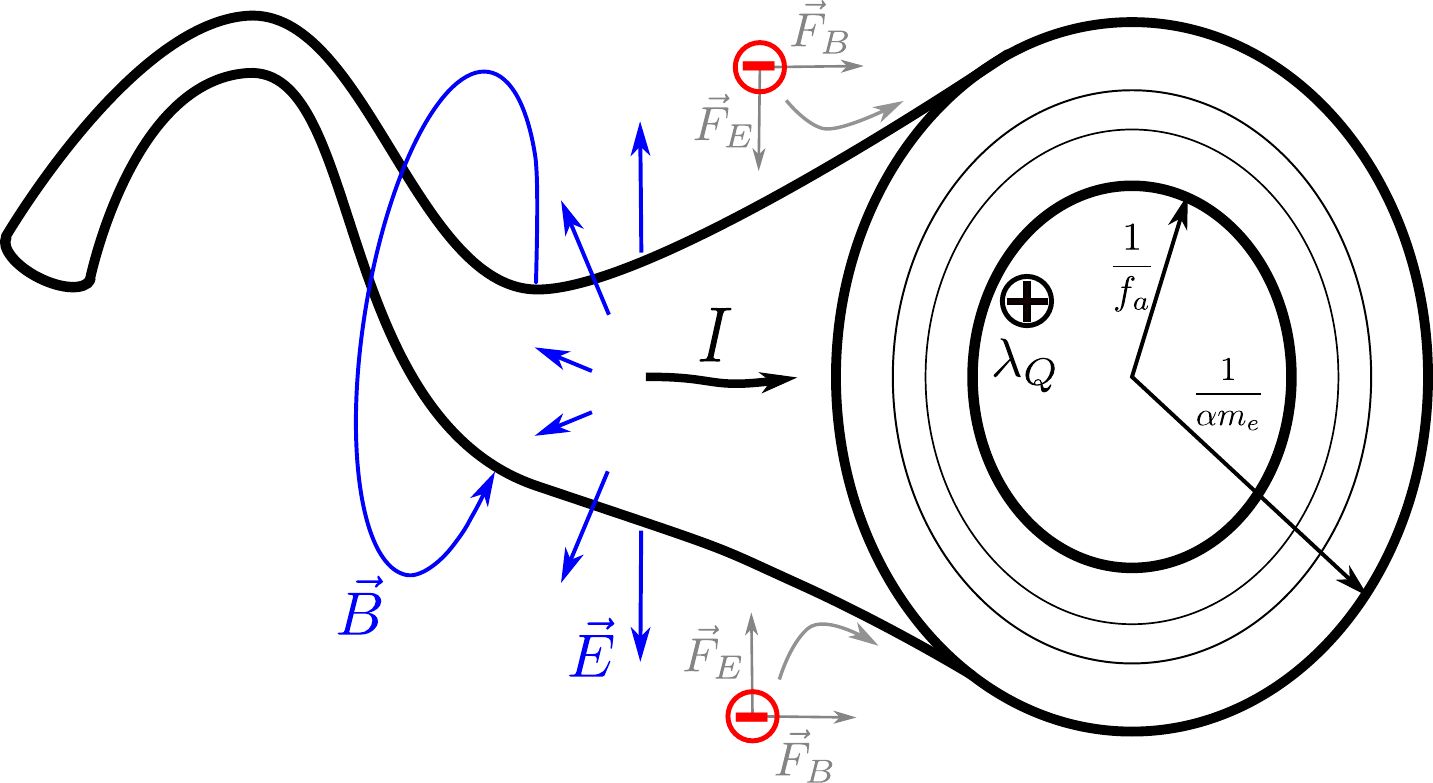}
\caption{A schematic figure of the cross section of the axion string. The string carries a positive charge density $\lambda_Q$ up to $10^9\,{\rm GeV}$ and a current in the direction of the black arrow. The charge and current on the string core create a strong electromagnetic field around the string core in the direction shown by the blue arrows. The strong electromagnetic field produces charged standard model particles and attracts the negatively charged SM particles into bound states with the string core, traveling in the same direction as the PQ fermions on the positively charged string core. The positively charged string core is surrounded by multiple layers of standard model particles from gauge bosons, pions to standard model quarks and leptons, forming a one-dimensional ``atom'' with radius of $1/\alpha m_e$.}\label{fig:stringcs}
\end{figure}

In this subsection, we discuss the structure surrounding a charged axion string as shown in figure~\ref{fig:stringcs}.  
The charged string obtains a large charge density and current $\lambda_Q$ because
the zero mode has been filled into a Fermi-sea on the string with Fermi momentum $p_F \sim \lambda_Q$.
From the point of view of the transverse plane, the zero mode is a bound state
with zero mass localized to a distance $1/m_{\rm PQ}$ of the string, where $m_{\rm PQ}$ is the bulk mass of the string localized fermion. The higher angular
momentum PQ bound states have energy $\sim m_{\rm PQ} > p_F$ and are thus not filled.

Much of the structure of a charged string can be determined from the
bound state properties, Eq.~\ref{Eq: bound states}.  The most
important point is to notice that there are in principle many bound
states with large negative energy, since there can be many different
momentum states along the string direction.
These bound state particles could also have $p_F \sim \lambda_Q$
cancelling most of the charge and would move down the string at close
to the speed of light.  This bound state would cancel much of the
electric field and magnetic field outside of the string core.

To cancel most of the charge of the string localized zero mode, at
leading order, you only need a single bound state (the $n=1$ state) as
the momentum in the direction of the string provides an extra quantum
number to fill.  
However, this is not the lowest energy configuration when we consider
electric charge repulsion between particles outside the string core. Instead, as soon as
most of the charge has been screened, the size of the bound state
orbit increases.  When considering high Z atoms, the outermost
electrons only see the effective charge of the nucleus screened by the
inner electrons.  Similarly, in our case, the outermost bound state
charged SM particles will also only see the effective charge density
$\lambda_Q^{\rm eff}(r)$ inside.  In the end, we find that the
effective charge as a function of radius ($r$) is
\bea
\lambda_Q^{\rm eff}(r) \sim \frac{1}{r}, \qquad \text{if} \qquad  \lambda_Q^{\rm eff}(r) \gg m_{\rm SM}.
\eea
Using this, we find that the charge density of bound state particles scales as $n_{\rm SM} \sim 1/r^3$ starting from $n_{\rm SM}\sim \lambda_Q^3$ close to the surface of the axion string, while the electric field falls as $1/r^2$ from the surface as a result.
The bound states are filled with all charged particles with $m_{\rm
SM} < \lambda_Q$.  Depending on how fast these states cool, bosons
such as the W boson in the inner orbits and pions on the outer orbits
may be significantly more occupied than the fermions.

The qualitative behavior changes when the bound state becomes non-relativistic, $\lambda_Q^{\rm eff}(r) \sim m_{\rm SM}$.  At this point, the bound state has positive energy but can still be filled efficiently by the Blandford-Znajek effect.
Additionally, because the radius scales as $r \sim 1/\sqrt{\lambda_Q^{\rm eff}(r) m_{\rm SM}}$, non-relativistic bound states can be filled more densely than relativistic bound states.  However, most heavy charged particles are not
stable and decay so that they can no longer be used to screen the charge beyond 
$r \sim 1/m_{\rm SM}^i \sim 1/\lambda_Q^{\rm eff}(r)$.
For example, the charged pions decay into leptons in vacuum. In the bound states structure outside the string, however, they can avoid decaying as long as all of the leptons the pions decay into have a Fermi momentum that is larger than the mass of the pion.  An analogous effect is what prevents neutrons from decaying inside of a neutron star.  Since $n_{\rm SM} \sim 1/r^3$ for any SM particles with $m_{\rm SM} < \lambda_Q^{\rm eff}(r)$, the pion decay is no longer Pauli blocked roughly $1/m_{\pi}$ away from the surface of the string core.

The scaling behavior of $\lambda_Q^{\rm eff}(r)$ finally changes when the lightest charged particle, the electron, becomes non-relativistic.  In this case the non-relativistic scaling of the radius gives
\bea
\lambda_Q^{\rm eff}(r) \sim \frac{1}{r^2 m_e}, \qquad  \text{if} \qquad \lambda_Q \ll m_e ,
\eea
where $m_e$ is the mass of the electron. 
Shortly after this, the charge per unit length becomes smaller than $\lambda_Q^{\rm eff}(r)\lesssim \alpha m_e$
and the charge density no longer looks continuous but instead looks point like from the point of view of the bound state electrons. As a result, we expect the whole charged particle cloud around the axion string core to have a radius of order $1/\alpha m_e$, in accordance to the well known fact that the size of all the atoms in the periodic table are of order $1/\alpha m_e$ independent of the charge of the nucleus at the leading order. Additionally, this layer of non-relativistic states are moving slower than the speed of light and can no longer follow the string core localized charges as they move along the string at close to the speed of light meaning that a small fraction ($m_e/\lambda_Q$) of the front end of the line of charges will be shedding the bound states screening the charge.
The last small fraction of charge density is likely screened by a thermal or non-thermal bath of particles surrounding the string rather than by bound states.

\section{Present day observable signatures}\label{sec:kamehamehas}

In the previous sections, we described a static axion string in a uniform magnetic field, and how a hyper-light axion string (axiverse string) will become charged and subsequently neutralized by the surrounding medium. In this section, we will discuss present day signatures of strings much longer than the coherence length of any magnetic field they encounter.
Passing through magnetic fields charges up a section of the string and surrounds it with a dense plasma
that all together travels along the string before colliding with another clump of charge yielding spectacular signatures.  These signatures would be observable with next generation radio telescopes.
While we discuss axion strings, a few of the signatures under consideration also occur for Witten's superconducting string~\cite{Witten:1984eb}.

\subsection{Charging up the string}\label{sec:chargeup}

As discussed in Sec.~\ref{sec:naculich} and Sec.~\ref{sec:galacticB}, as a string passes through a magnetic field, it will acquire a very high charge density. There are two figures of merit for determining how good a region of space with magnetic field $B$ and radius $d$ is at charging up the string: $\lambda_Q\sim \alpha B d v$, which determines the average charge density reached and $Q_{\rm string}\sim \alpha B d^2$, which determines how much charge can move onto the string. In the following, we show some of the typical magnetic field regions and how much they can charge up a string.

\paragraph{Galactic B-fields}  To characterize the effect on an axion string when it passes through a galaxy, we take the magnetic field of the Milky Way as an example. The galactic magnetic field of the Milky Way has a size of order $5 \,{\rm \mu G}$ with a coherence length of order $10\,{\rm kpc}$ inside the plane with a height of order $3\,{\rm kpc}$~\cite{2019MNRAS.484.3646S}. The charge density on the string reaches a value of order
\begin{equation}
\lambda_Q = \frac{e^2 \mathcal{A}}{2\pi} B_{\rm galaxy} d_{\rm galaxy} v_s \approx 3 \times 10^8 \,{\rm GeV} \left(\frac{\mathcal{A}}{1}\right)\left(\frac{B_{\rm galaxy}}{5\,{\rm \mu G}}\right)\left(\frac{v_s}{0.1}\right)\left(\frac{d_{\rm galaxy}}{10 \,{\rm kpc}}\right)
\end{equation}
whereas the total charge that is carried by such a string section can be as large as
\begin{equation}
Q_{\rm string,\,\,galaxy} \simeq \frac{e^2 \mathcal{A}}{2\pi} B_{\rm galaxy} A_{\rm galaxy} \approx 10^{45} \left(\frac{B_{\rm galaxy}}{5 \,{\rm \mu G}} \right)\left( \frac{A_{\rm galaxy}}{100 \,{\rm kpc^2}} \right)
\end{equation}
Here, we take the B-field in the galaxy to be mostly parallel to the disk of the galaxy and the cross section area of the magnetic field region $A_{\rm galaxy}$ to be the diameter times the height of the B-field region.  

The rate at which strings pass through galaxies can be estimated to be 
\begin{equation}
\label{eq:galaxyrate}
\Gamma \simeq \xi H^3 N_{\rm galaxy}  L_{\rm string} d_{\rm galaxy} v_s \approx 2\times10^{-4} \,{\rm yr^{-1}} \left(\frac{\xi}{10}\right)\left(\frac{N_{\rm galaxy}}{10^{12}}\right) \left(\frac{d_{\rm galaxy}}{10 \,{\rm kpc}}\right)\left(\frac{v_s}{0.1}\right) 
\end{equation}
where $\xi$ is the number of strings per Hubble patch 
and each of these crossings last $t = d_{\rm galaxy}/v_s \approx 3\times 10^5 \,{\rm yrs}$. This means that there are $\mathcal{O}(100)$ galaxy - string crossings happening at any given time.
Similar to galaxies, the string can also come across magnetic field domains with similar properties in galaxy clusters, each with their own orientations and continuously charging up and discharging the string.

\paragraph{Intergalactic B-fields} Another situation in which charge can appear on the string comes from the interaction of the string with intergalactic B-fields. The intergalactic medium has a magnetic field with strength of order $1 \sim 10 \,{\rm nG}$ and a coherence length of order $0.1 \,{\rm Mpc}$. This magnetic field can charge the string up to  a charge density of 
\begin{equation}
\lambda_Q \approx \frac{e^2 \mathcal{A}}{2\pi} B_{\rm ig} d_{\rm ig}  v_s \approx 10^6 \,{\rm GeV} \left(\frac{\mathcal{A}}{1}\right)\left(\frac{B_{\rm ig}}{{\rm n G}}\right)\left(\frac{d_{\rm ig}}{0.1 \,{\rm Mpc}}\right) \left(\frac{v_s}{0.1}\right) .
\end{equation}
Each string segment of $0.1 \,{\rm Mpc}$ length contains a total charge of $10^{44}$ stored on the string core.

\paragraph{Magnetars and pulsars} Magnetars carry the strongest magnetic fields in the universe, with field strengths up to $10^{15} \,{\rm G}$~\cite{Harding:2006qn}.  Magnetars can charge the string to an extremely large charge density $\lambda_Q \sim 10^{12} \,{\rm GeV}$ locally, which could lead to releasing PQ fermions into the neutron star environment when $\lambda_Q \gtrsim m_{PQ}$. However, such a density is only on a relatively small section of the axiverse string, and the total charge on the string from such a crossing is $Q_{\rm string}
 = 10^{32}$. 
 
Unfortunately, the chance that a string passes near a magnetar/pulsar is close to zero. The rate of such events is
\begin{equation}
\Gamma = \xi n_{\rm pulsar}  (L_{\rm string} d_{\rm NS}) v_s \approx 
10^{-17}/ \,{\rm yr} \left(\frac{\xi}{10}\right)\left(\frac{d_{\rm NS}}{10\,{\rm km}}\right)\left(\frac{N_{\rm pulsar}}{10^{16}}\right) \left(\frac{v_s}{0.1}\right) 
\end{equation}
where $N_{\rm pulsar}$ is an estimate of the total pulsars in the universe. As it is evident, such a scattering is unlikely to happen even once in the lifetime of the universe.

\paragraph{Stellar B-field} Stars can carry a magnetic field that is of order a few gauss over a size of a few light seconds. A star is unlikely to significantly affect a string that passes through with speed of order $0.1$. However, as we will show later, the string can potentially affect the star it passes by if the string is trapped by the galaxy and moving very slowly.

In summary, the different sections of a long string can obtain a huge charge as they pass through  galaxies and galaxy clusters. However, this charge is not observable by a distant observer as the string is almost always overall neutral. Just to be clear, we list again the four main reasons. Firstly and most importantly, as discussed in section~\ref{Sec: gmt}, the charge is locally shielded by the dense SM plasma around the string core. Secondly, the orientation of the magnetic field regions in the universe is totally random.  Thirdly, the dipolar nature of the galactic B-field makes sure that as the string goes through the whole galaxy, it gets zero overall charge. And lastly, the charge stored on the string core plus the charge in the axion profile around the string core sums up to a constant that can only change in rare situations (see appendix~\ref{sec:earlyuniverseB}).

\subsection{Propagation along the string}

In order to eventually understand collisions, we need to first
understand how the charge overdensities disperse on the string. 
As the string moves through the magnetic field, there is an average
charge density $\lambda_Q \approx 2\alpha B L v_s$ being generated
locally and is constantly dispersing at the speed of light in both
directions. The phenomenon of the dispersion of charge overdensities at
close to the speed of light is quite independent of the properties of
the charge
carriers on the string. No matter which direction the ``$\pm$'' charged
charge carriers move, the current can always point in both
directions.
Thus when the string has finished moving through the uniform magnetic
field patch,  the average charge density becomes $\lambda_Q \sim
2\alpha \Phi/t$. 

As mentioned in Sec.~\ref{sec:galacticB}, the charge on the string
core quickly becomes surrounded by a bound state plasma.  Thus the
charge overdensities on the string core when viewed from the outside
are almost completely shielded.
These charged particles are very tightly bound to the string core, and
as the charge overdensities move along the string, these tightly bound
states will move in the same direction as the current, forming an
extremely dense and high energy plasma that is traveling at nearly the
speed of light together with the overdensity on the string.  

To understand the qualitative behaviors of this charge dispersion
along the axion string, it is helpful to zoom out and think about a
generic superconducting loop. A superconducting loop with no
resistance does not dissipate, and the charge overdensities on the
string can in principle oscillate forever, instead of reaching the
lowest energy configuration where the charge is uniform on the string.
The dissipation on the string actually comes entirely from the bound
state Standard Model particles around the string core, which disperse
together with the charge overdensities on the string.

To see this, let us consider a tightly bound cloud of Standard Model
particles that is lagging behind the charges localized on the string
core as they disperse. The lagging SM particles around the string core
create an electric field which slows down the dispersing charge on the
string core. For zero mode fermions on the string, the electric field
decreases their momentum (Fermi momentum), which by momentum
conservation means that the SM bound states outside obtain a larger
momentum. During this process, the charge density on the string core
$\lambda_Q$ decreases as the Fermi momentum decreases while the total
charge on the string core is conserved. The SM particles outside of
the string experience a decrease in the electric field and rearrange
into the new eigenstates of the changing potential, emitting radiation
in the process. This process allows energy to be released by our
one-dimensional ``atom'' and is similar in spirit to the $\gamma$
photons emitted during the rearrangement of an electron cloud
following an $\alpha$ and $\beta$-decay in atomic physics. This
process, together with the process we discuss in the next subsection,
allows the system we consider to dissipate energy and reach its ground
state.

The strong magnetic fields in the universe are usually dipole like and
as a result, it is expected that as the string moves past a galaxy, it
will eventually encounter a patch of magnetic field that has the
opposite orientation and the charge eventually averages to zero.  We
expect the charge overdensities on the string to display a profile
that resembles a wave packet, before colliding with the wave packet
coming from a nearby galaxy or another magnetic field patch. These
collisions will be the topic of the rest of the section.

\subsection{Collisions on the string}

\begin{figure}
\centering
\includegraphics[width=0.8\textwidth]{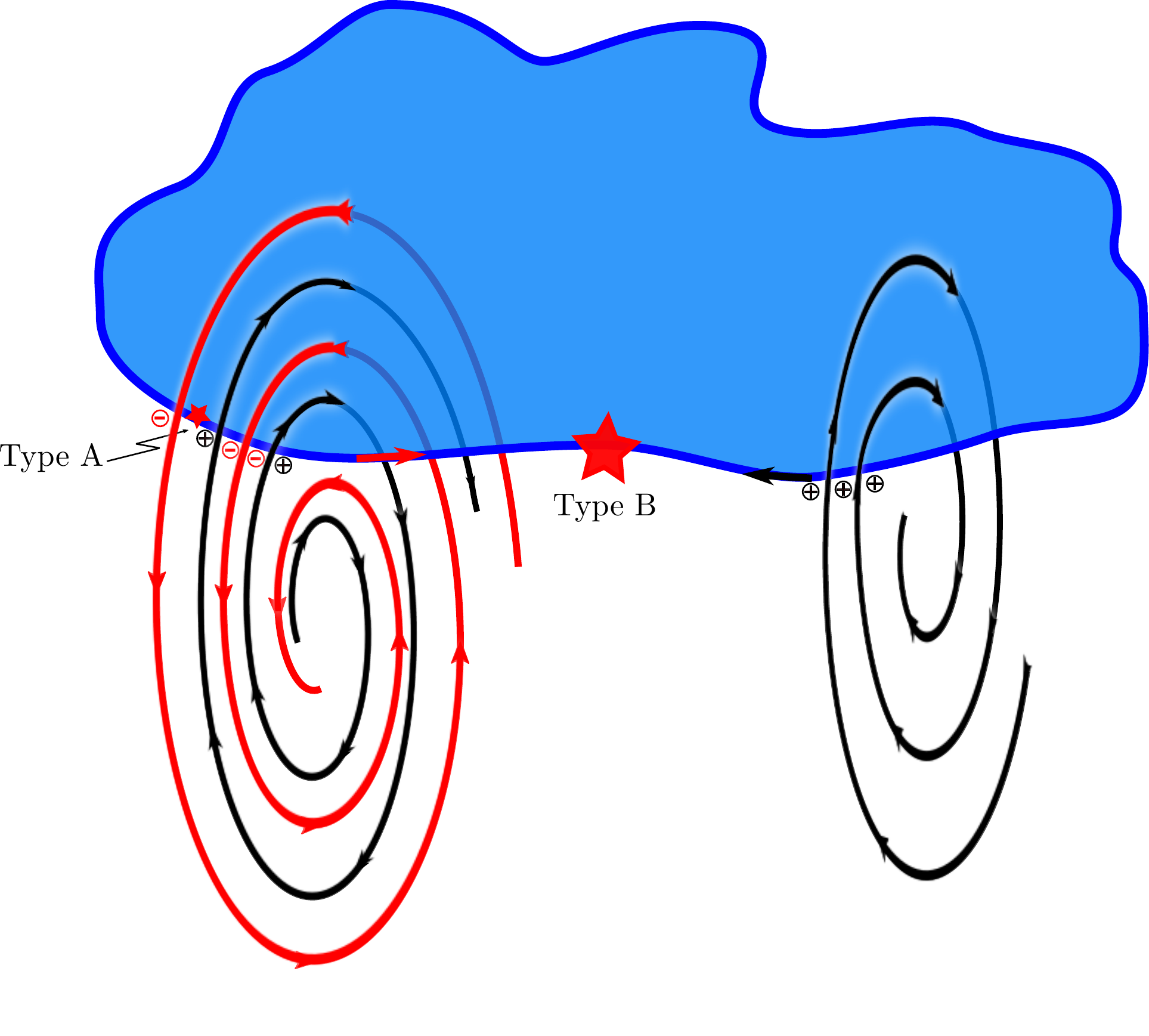}
\caption{A schematic presentation of the two types of plasma collisions that can occur in the universe, as the string is moving through the universe. Type A collisions (small red star) happen when the axion string (blue solid line) moves through magnetic field regions with different orientation in the dipolar galactic magnetic field~\cite{2015A&ARv4B}. This leads to a collisions inside the galaxy. Type B collisions (big red star) happen when the axiverse string moves through different magnetic field regions of different galaxies inside a galaxy cluster, or different magnetic field patches inside the galaxy cluster, which leads to collisions outside of a galaxy. Both the collision and the charging and discharging of the string provides us with observable signals.}\label{fig:kamehameha}
\end{figure}

Violent collision events can occur if nearby string segments go through magnetic field regions with opposite orientations. As the charge overdensities travel along the string, the SM charged particle cloud outside the string will move in tandem with the current on the string.
As the strings go through various magnetic field regions, different sized charge overdensities are formed and collide. For a visual representation of this collision, click \href{https://youtu.be/Ilzm8K-AfGs?t=110}{here}. 
The main observable effects of the strings are (see figure~\ref{fig:kamehameha}):
\begin{itemize}
\item Emission from the collision
\item Emission from the surface of the plasma as the string charges and discharges
\item Stellar objects the string goes through 
\end{itemize}

We will first discuss the most spectacular of all the observable effects, the collision. The number of encounters between galaxies and strings, $N_K$, in the universe depends on the number density of strings, $n_s \sim \xi H^3$, where $\xi$ is a free parameter that can in principle be fixed by simulations~\cite{Gorghetto:2018myk,Gorghetto:2020qws,Buschmann:2019icd}. 
\begin{equation}
N_K \simeq \xi H^3 L_{\rm string} N_{\rm galaxy} A_{\rm galaxy}\approx 100 \left(\frac{\xi}{10}\right)\left(\frac{N_{\rm galaxy}}{10^{12}}\right) \left(\frac{A_{\rm galaxy}}{(10 \,{\rm kpc})^2}\right),
\end{equation}
where $N_{\rm galaxy}$ is the total number of galaxies in the universe and $A_{\rm galaxy}$ is the cross section area of a typical galaxy. 
For simplicity we take the total number of galaxies to serve as an estimate of the number of strong magnetic field regions in the universe.
The string length $L_{\rm string} \sim 1/H$ is the typical length of a string in the universe today. 

As mentioned in the previous subsection, as the charge density moves, it is also losing energy and spreading out.  In order to have a collision, two magnetic field regions of opposite charge need to be close by.  A simple scenario where this occurs is if two galaxies are nearby and the string is going through them both.  Then the charge overdensity created by one can quickly propagate to the next galaxy and cause a collision before a significant amount of energy is lost.

Out of these $\mathcal{O}(100)$ galaxies that the strings go through, many are embedded inside of a galaxy cluster.  Galaxy clusters host $N_{\rm gc} = \mathcal{O}(10^2 \sim 10^3)$ galaxies within a $\mathcal{O}(10 \,{\rm Mpc})$ region.  
If the string passes through two galaxies in the galaxy cluster before the charge overdensity of the first has had time to leave that section of the string, then a collision will occur.
The probability that a collision occurs because of this is 
\begin{equation}
 p \approx \frac{p_0}{2}  n_{\rm galaxy} (d_{\rm cluster} d_{\rm galaxy}) v_s  d_{\rm cluster} \approx  0.1 \left(\frac{v_s}{0.1}\right) \left(\frac{p_0 N_{\rm gc}}{10^3}\right) \left(\frac{d_{\rm galaxy}}{10\,{\rm kpc} }\right) \left(\frac{10\,{\rm Mpc}}{d_{\rm cluster}}\right),
\end{equation}
where $p_0$ is the percentage of galaxies in the universe that belongs to a galaxy cluster, and $v_s$ is the velocity of the string. This suggests that there can be $p N_K \sim 10$ of these collisions happening at any given time in the observable universe. 

Depending on how quickly the charge overdensities spread out, these collisions take anywhere between $t \sim 10$ kpc $\sim 10^4$ years to $t \sim 10$ Mpc $\sim 10^7$ years to complete.
In these collisions, the bound state particles colliding release energy of order their binding energy
\begin{equation}
E \simeq \frac{e \lambda_Q}{2 \pi}  \log \left(f_a L\right) .
\end{equation}
Energy is released at a rate of
\begin{equation}
P \simeq \frac{\lambda_Q^2}{2\pi} \log \left(f_a L\right) \approx 10^{40} \,{\rm erg/s} \left(\frac{\lambda_Q}{10^9 \,{\rm GeV}}\right)^2,
\end{equation}
which can be up to seven orders of magnitude brighter than the Sun. 

The observability of such a signal depends on how the energy is emitted.
Unfortunately, the collisions we are describing are not like any other sources of radiation known to exist in the universe, so it would likely require simulations to determine the properties of the collision point.  
The dense structure at and near the collision point could allow the production of many different types of observable signals from very high energy cosmic rays, neutrinos, $\gamma$-rays and hard X-rays to lower energy emissions such as optical and radio frequency photons. 
In the following, we discuss various possibilities qualitatively.

The collision of the bound state plasma can produce high energy neutral particles that escape the dense medium.  These high energy cosmic ray emissions can have energies that are comparable to or even larger than the TeV scale.  Significant neutrino and gamma ray emission is plausibly emitted from the surface of the region where the pion density is very high, at a radius $r \sim 1/m_{\pi}$ from the string core. The charged and neutral pions produced in the collision lead to neutrinos and $\gamma$-ray photons from their decay.  The energy carried per neutrino and photon is possibly much larger than $m_{\pi}$ due to acceleration in the electromagnetic field around the string like what happens in an astrophysical jet, though whether such a jet like structure would eventually develop at the plasma collision point is unclear.

Apart from high energy emission and kink formation, it is conceivable that the heated up region near the collision point can also emit thermal photons from the surface of the hot plasma around the collision point. The two charge overdensities that collide will likely have $\mathcal{O}(1)$ different densities since the different magnetic field patches of different galaxies are likely $\mathcal{O}(1)$ different and time dependent. This leads to two main consequences. Firstly, this time dependence suggests that the collision point can in principle be moving back and forth. However, over the observational time window of any realistic experiment, such a movement would be too tiny and one can treat the collision point as fixed for all observational purposes. Secondly, the asymmetry between the colliding overdensities means that there will be some extra charged matter that starts to pile up around the point of collision. Over the course of the duration of the collision, this excess energy density can pile up to a significant fraction of a solar mass. It is possible that all of this matter can form a giant fireball, which emits mainly thermally or non-thermally at lower frequencies. 

The last plausible source of emission comes from the existence of a strong magnetic field around the string. Unlike the electric field, which is neutralized by the standard model plasma around the string, the magnetic field can in principle survive in much the same way magnetic fields survive around magnetars.  This strong magnetic field can trap the outgoing charged particles and cause them to emit synchrotron radiation. It is conceivable that synchrotron radiation is the dominant component of energy loss from the collision point in analogy to how pulsars lose most of their energy~\cite{Harding:2006qn}, as the magnetosphere surrounding a pulsar is somewhat similar to the matter surrounding the collision point.  
Moreover, since the magnetic field in our case comes from a lasting current and decays as $1/r$ in vacuum instead of the $1/r^3$ scaling of dipoles, radio emission can in principle be coming from a much larger volume around the collision point, which can result in an extremely bright source. The power and spectrum of emission into radio depends on both how much energy is being released by the collision, as well as the density profile of the plasma and the magnetic field profile inside the plasma away from the string core.  The latter of these is unfortunately unattainable with analytical estimates.  These strong magnetic fields can also be the basis of the jet-like structures mentioned earlier, which would lead to a source of ultra high energy cosmic rays.

\subsection{Observational prospects}\label{sec:observe}

The collision of charge overdensities produces an extremely bright point source of continuous emission. Such a source is unfortunately usually very far away since the expected distance of the Earth to the nearest string is $\mathcal{O}(1/H_0)$. This suggests that the collision can be modeled as a point source and its energy flux can be as large as
\begin{equation}
\label{Eq: signal brightness}
\frac{P}{A} \simeq 10^{-16} \,{\rm erg/s/cm^2}\left(\frac{\xi \mathcal{A}^2}{1}\right).
\end{equation}
given the constraint from CMB measurements~\cite{Agrawal:2019lkr}. 

The best window we have to look for such an event is with a radio telescope, for example the Green Bank Telescope (GBT), Arecibo, and the Five-hundred-meter Aperture Spherical radio Telescope (FAST). The Five-hundred-meter Aperture Spherical radio Telescope (FAST) and Square Kilometer Array (SKA) is suitable for a dedicated measurement while Canadian Hydrogen Intensity Mapping Experiment (CHIME) and Low-Frequency Array (LOFAR) are suitable for a survey.

In the following, we will use CHIME as a benchmark to show the observational prospects of a wide survey~\cite{10.1117/12.2056962,Amiri:2017qtx}, and FAST and SKA for the prospects of a dedicated measurement~\cite{7303195}. Over the time span of an experiment, the collision point can at most move by $\mathcal{O}({\rm pc})$, which would correspond to a change of $10^{-4}$ arcseconds for a string that is of order ${\rm Gpc}$ away.  This change is much smaller than the angular resolution of all present and future radio surveys and the object can be well approximated as a stationary point source. For such a source, the sensitivity is determined by the system-equivalent flux density (SEFD) of the telescope, the bandwidth $B$ of the signal and the integration time $t_{\rm int}$. 
\begin{align}
\left.\frac{P}{A}\right|_{\rm exp} &= {\rm SEFD} \left(\frac{B}{t_{\rm int}}\right)^{1/2} \nonumber\\
&= 2 \times 10^{-18} {\rm erg/s/cm^2} \left(\frac{\rm SEFD}{10^4\,{\rm Jy}}\right) \left(\frac{B}{\rm GHz}\right)^{1/2}\left(\frac{1000\,{\rm hr}}{t_{\rm int}}\right)^{1/2} \, & {\rm (Survey)}\nonumber\\
&= 5 \times 10^{-20} {\rm erg/s/cm^2} \left(\frac{\rm SEFD}{10\,{\rm Jy}}\right) \left(\frac{B}{\rm GHz}\right)^{1/2}\left(\frac{\rm hr}{t_{\rm int}}\right)^{1/2} \,& {\rm (Dedicated)} 
\end{align}
Comparing these estimates with Eq.~\ref{Eq: signal brightness}, we see that a survey will be able to find a signal that has roughly one percent of the total emission in radio frequencies. The exact spectrum coming from such a collision is unknown without a dedicated simulation, but the prospect is very encouraging.  
Things improve with a dedicated search, which will be able to discover such a source even if only $10^{-4}$ of the energy falls into the frequency range of order 1 to 10 GHz in an experiment like FAST or SKA~\cite{7303195}. 

A dedicated search will be unable to find the collision without
knowing where to look ahead of time because of the small field of
view.  However, the same strings we are studying also lead to
observable CMB polarization rotations that can be looked for both in
Fourier space~\cite{Agrawal:2019lkr} and also in position/angular
space~\cite{LiangDai}. These searches, especially edge detection
methods, give us the ability to locate a long string in the sky to
very good precision.  We would thus be able to know exactly which
galaxies these long strings could be going through and where the
collisions could occur. The $\mathcal{O}({\rm deg}^2)$ field of view
of FAST and SKA allows us to have the whole galaxy cluster at cosmological
distances inside the field of view of the telescope, while the
$\mathcal{O}({\rm arcsec})$ angular resolution allows us to identify
the various galaxies inside of the cluster.
This angular resolution would allow us to focus on searching for collisions in the region of the galaxy clusters where galaxies are not present, something important for background rejection. 

If the energy being emitted is dominantly in optical frequencies, then the energy released by these collisions would be roughly one order of magnitude brighter than the faintest object that can be looked for with the Hubble Space Telescope and two orders of magnitude brighter than those that can be found by the future James Webb Space Telescope (JWST).  If the collision point emits dominantly at even higher energies, then the flux can potentially be detected by X-ray telescopes like XMM-Newton and Chandra but would likely be beyond the reach of Fermi if the emission is directionless. However, if jet like structures are created, then it is possible that we would get lucky and discover some of these collisions as a source of very high energy cosmic rays.

\subsection{Other observable effects}\label{sec:otherobs}

In much of this section, we focused on the case where $\lambda_Q \simeq 10^9 \,{\rm GeV} \ll f_a$, where the SM charged particles around the string can only take away a very small fraction of the total kinetic energy and momentum of the string. However, when $\lambda_Q > f_a$, the Lorentz force of $B I L$ and ejected charged particles can in principle significantly slow down the axiverse string and a section of the axiverse string can be trapped inside a galaxy. Over the age of the universe, the different string sections can pick up thousands of these galaxies, forming a ``necklace'' of galaxies attached to a single axiverse string. Such a scenario can be tested if we find the axiverse strings in the CMB measurements with edge detection methods~\cite{LiangDai}. Knowing the locations of the axiverse strings, we can ask if there is an overdensity of galaxies that is correlated with the location of long axiverse strings in the sky.

When the string is trapped inside the galaxy, the chances for it to hit a main sequence star also significantly increases. The rate of such a star crossing event per galaxy is
\begin{equation}
\Gamma \simeq n_{\rm star} (d_{\rm star} d_{\rm galaxy}) \max \{v_s, \,\,v_{\rm star}\} \approx 10^{-8}/{\rm yr} \left(\frac{N_{\rm star}}{10^{12}}\right)\left(\frac{d_{\rm star}}{\rm second}\right)\left(\frac{10\,{\rm kpc}}{d_{\rm galaxy}}\right)^2\left(\frac{v_{\rm star}}{10^{-3}}\right)
\end{equation}
Over the age of the universe, the axiverse strings in the universe could have been trapped by up to $10^6$ galaxies (see Eq.~\ref{eq:galaxyrate}) over the age of the universe. This means that these star crossing events happen roughly once every one hundred of years, and a total of $10^7 \sim 10^8$ stars have been impacted by the passing axiverse string. Similarly, the captured string can also encounter neutron stars with a rate of 
\begin{equation}
\Gamma \simeq n_{\rm NS}  (d_{\rm NS} d_{\rm galaxy}) \max \{v_s, \,\,v_{\rm NS}\} \approx 10^{-16}/{\rm yr} \left(\frac{N_{\rm NS}}{10^{8}}\right)\left(\frac{d_{\rm NS}}{10 \,{\rm km}}\right)\left(\frac{10\,{\rm kpc}}{d_{\rm galaxy}}\right)^2\left(\frac{v_{\rm NS}}{10^{-3}}\right)
\end{equation}
per galaxy, if we assume there are roughly $\sim 10^8$ neutron stars in a galaxy~\cite{Sartore:2009wn}. This means that there have been a few neutron stars that have been impacted by the passing axiverse string. This rate can significantly increase if a large fraction of these neutron stars have large enough magnetic field to slow down a small section of axiverse string significantly once it passes through the much larger magnetosphere of the neutron star (see appendix~\ref{sec:earlyuniverseB}). Naive expectation will suggest that these stars and neutron stars that are captured by the axiverse string will be destroyed by the occasional passing of plasma waves and explode, though no existing analysis of cosmic strings addresses this question. We leave this question to future work.

As has already been pointed out in Ref.~\cite{Witten:1984eb}, axion strings can also affect the magnetic field profiles inside a galaxy, providing additional contributions to the dynamo process. These effects could potentially enhance the signal that we can look for. We leave these interesting questions to future work.

\begin{figure}
\centering
\includegraphics[width=0.8\textwidth]{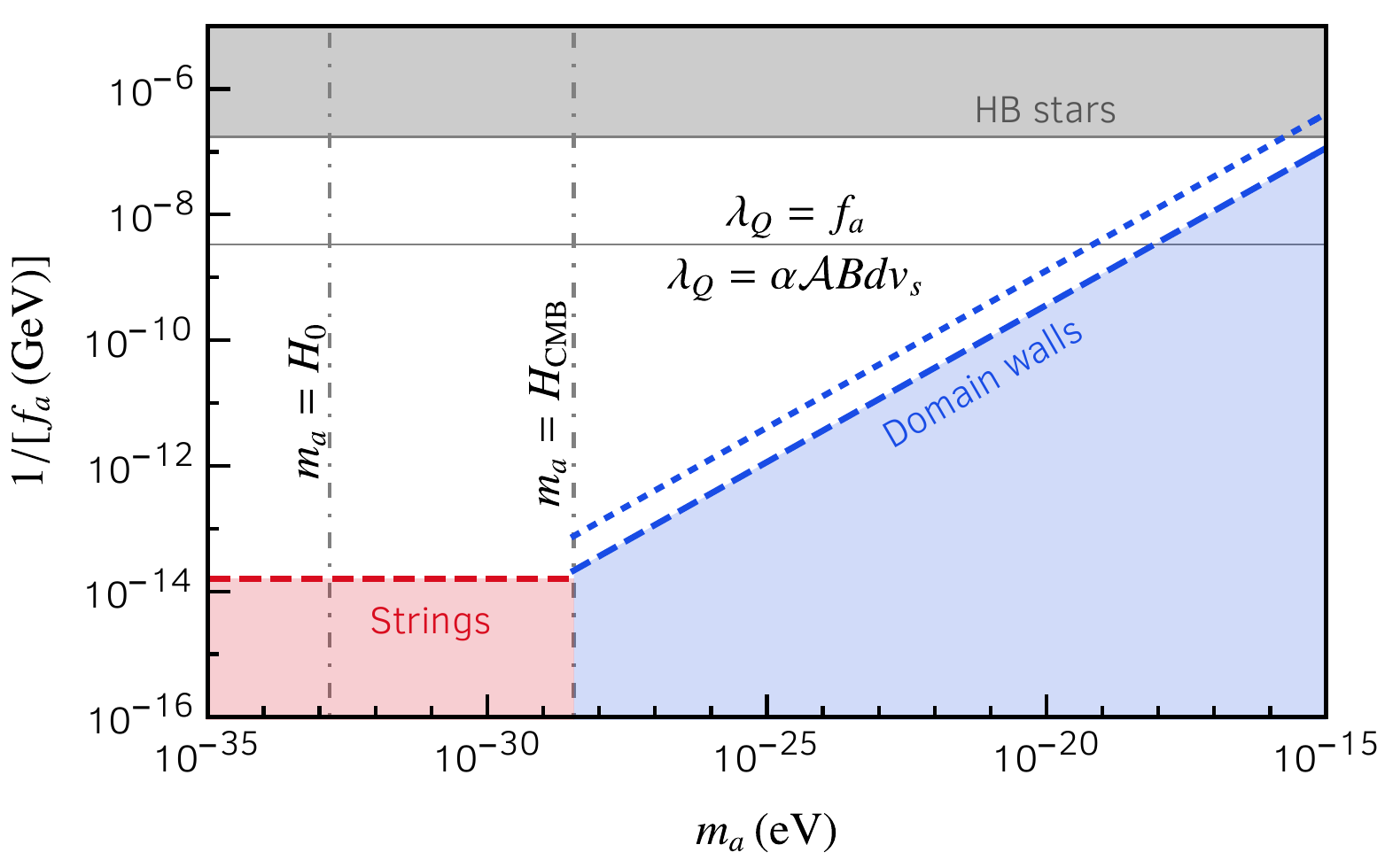}
\caption{The open parameter space of hyperlight axions. The gray shaded
region is excluded by cooling bounds coming from Horizontal Branch stars~\cite{Ayala:2014pea} (assuming $\mathcal{A} = 1$). The
red shaded region is excluded by CMB measurements of strings coming from their gravitational effects~\cite{Charnock:2016nzm}.
The blue region is excluded when $N_{\rm DW} \neq 1$ due to the temperature fluctuations induced by the gravitational
potential of the domain wall~\cite{Sousa:2015cqa}. Dashed and dotted blue and red lines correspond to $\xi = 10$ and $100$, respectively. 
The horizontal gray line divides the parameter space into two parts. In the lower region, the charge density is always smaller than $f_a$, and the charge density is set by the magnetic field properties in the galaxy (see Sec.~\ref{sec:observe}). In the upper region, the charge density $\lambda_Q$ on the string core can reach and saturate at $f_a$ (assuming $m_{PQ} = f_a$), and the string can potentially be trapped by a galaxy (see Sec.~\ref{sec:otherobs}).
The signals discussed in this paper apply to $\xi \geq 1$ in all of the allowed parameter space in
this plot.}\label{fig:massfa}
\end{figure}

\section{Conclusions}\label{sec:conclusion}

In this paper, we studied in detail some observable consequences  of axion strings coming from the effect electromagnetic fields have on an axion string as the they pass by the string. The passing magnetic field moves charge from the axion profile around the string to the superconducting axion string core. This charge movement is a consequence of the beautiful physics of anomaly inflow~\cite{Callan:1984sa,Bardeen:1984pm,Harvey:2000yg} and the superconductivity of axion strings has been explored by seminal papers in the 1980s~\cite{Naculich:1987ci,Kaplan:1987kh,Manohar:1988gv,Harvey:1988in}. Whereas it is unlikely that these effects significantly affect the evolution of QCD axion strings, for axion strings of hyperlight axions (axiverse strings) that can persist until today, these effects can lead to remarkable signals that we can look for.

Axiverse strings passing through galaxies obtain a charge on their core stored in the form of chiral zero modes and the profile of the radial and angular modes.  In a typical galaxy, the charge density along the string $\lambda_Q$ can become as large as $\min \{m_{PQ}, 10^9 \,\rm GeV\}$ (see Fig.~\ref{fig:massfa}), with a total charge as large as $10^{45}$. This huge charge density gives rise to large electromagnetic fields around the string core, leading to swift particle production of bound state Standard Model charged particles. This suggest the produced charge overdensities on the axion string core are neutralized by a dense cloud of bound Standard Model plasma around the string, forming a one dimensional atom. 

The zero modes on the axion string, as well as the dense plasma outside, travel at the speed of light along the string. These packets of high energy plasma collide at a center of mass energy of up to $10^9$ GeV. These collisions create a giant fireball of particles in a strong magnetic field, an environment ideal for forming a strong source of radio emission. These bright sources can have luminosities up to seven orders of magnitude larger than the solar luminosity and last for thousands to even millions of years, which makes them visible to radio telescopes even when at cosmological distances.

\begin{figure}
\centering
\includegraphics[width=0.8\textwidth]{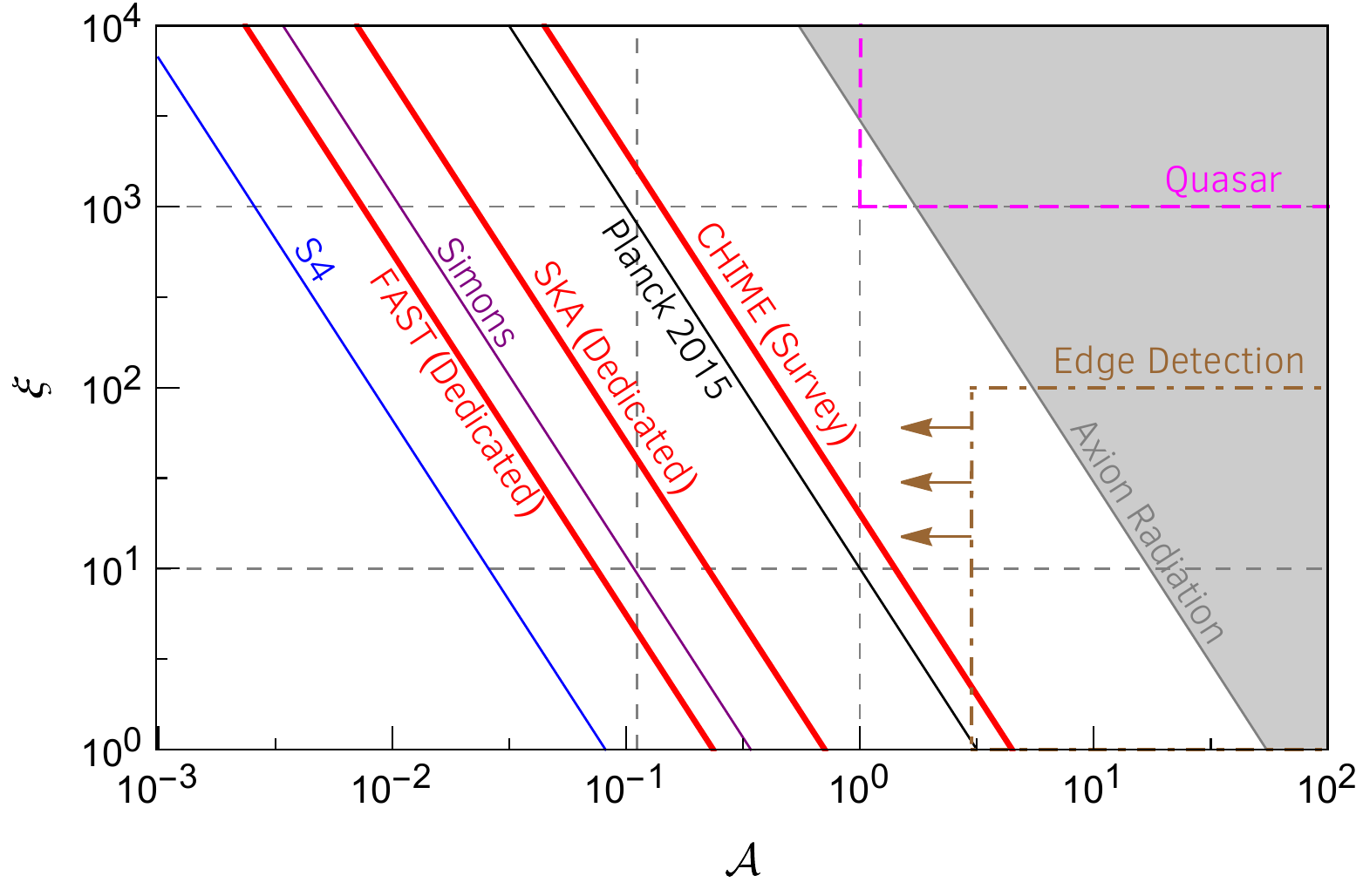}
\caption{We show the sensitivity of current and future experiments to the two independent parameters of
an axion string network with a photon coupling; the number of axion string per Hubble volume $\xi$ and the
strength of the axion string-photon coupling $\mathcal{A}$.  
The thick red lines show the sensitivity of a survey (CHIME) and a dedicated observation (FAST and SKA) to the radio signal coming from collisions on the string.
Both lines assume that $10^{-3}$ of the total emitted energy is in the sensitivity band of the detector. 
The strings can also be looked for in the CMB due to polarization rotation effects~\cite{Agrawal:2019lkr}.
The black solid line is the estimated
current sensitivity of Planck while the purple and blue lines show the sensitivity of future CMB experiments.
The shaded grey region is excluded due to emission of axion radiation and its effect on the CMB.
The brown dot-dashed line shows the prospects of edge detection methods with experiments similar to the Simons observatory (very qualitative). The gray dashed gridlines mark the regions of parameter space
of particular theoretical interest.}\label{fig:money}
\end{figure}

These radio sources can be looked for with radio surveys like CHIME as well as with dedicated radio measurements with, for example FAST and SKA.  There is a very non-trivial cross correlation one can perform between the edge detection of axiverse strings with CMB measurements~\cite{Agrawal:2019lkr,LiangDai} and the radio sources we look for in this paper (see Fig.~\ref{fig:money} for a summary).  The same axiverse string that gives rise to the polarization rotation signal in the CMB also gives rise to the bright radio sources we look for in this paper due to the same axion photon coupling.  If a string were found in the CMB, we could point a radio telescope at the exact galaxy clusters that the string passes through and do a dedicated measurement. We summarize the observational prospects of these various searches in Fig.~\ref{fig:money} along with the existing constraints obtained previously in Ref.~\cite{Agrawal:2019lkr}.
Seeing the same axion string in multiple ways would be astounding.

\section*{Acknowledgement}
We thank Robert Lasenby and Davide Racco for useful discussions and comments on the draft. 
The authors also thank 
Asimina Arvanitaki,
Liang Dai,
Neal Dalal, 
Daniel Ega\~na-Ugrinovic,
Shamit Kachru,
John March-Russell, 
Julian Mu\~noz,
Rob Myers,
Ue-Li Pen,
Raman Sundrum
and Ken Van Tilburg 
for useful conversations. We also thank Akira Toriyama for inspiration. 
The authors acknowledge the KITP for its hospitality during the inception of this project and National Science Foundation under Grant No. NSF PHY-1748958. 
PA is supported by STFC under Grant No. ST/T000864/1.
AH and GMT are supported in part by the NSF under Grant No. PHY-1914480 and by the Maryland Center for Fundamental Physics (MCFP). 
Research at Perimeter Institute is supported
by the Government of Canada through Industry Canada
and by the Province of Ontario through the Ministry of
Economic Development \& Innovation.

\appendix

\section{Charge (non) conservation in the axion string system}\label{sec:earlyuniverseB}

In this section, we demonstrate how the total charge on the axion string plus the Goldstone Wilczek charge off of the string can change (see Eq.~\ref{Eq: sum}).
In any interaction between the string and a magnetic field, the total charge does not change.
However, zero-mode fermions can go onto and off of the string in many other ways and thus
change the total charge of the string - axion system $Q_{\rm tot}$.
Aside from the zero-mode fermions, the charge held in the current $j_A$ can also leave the string by emitting PQ fermions
as axion field configurations carry both electric charge and PQ fermion number.
In this appendix, we discuss several different types of processes that can change the total charge of the string.

\subsection{String crossing}

\begin{figure}
\centering
\includegraphics[width=0.5\textwidth]{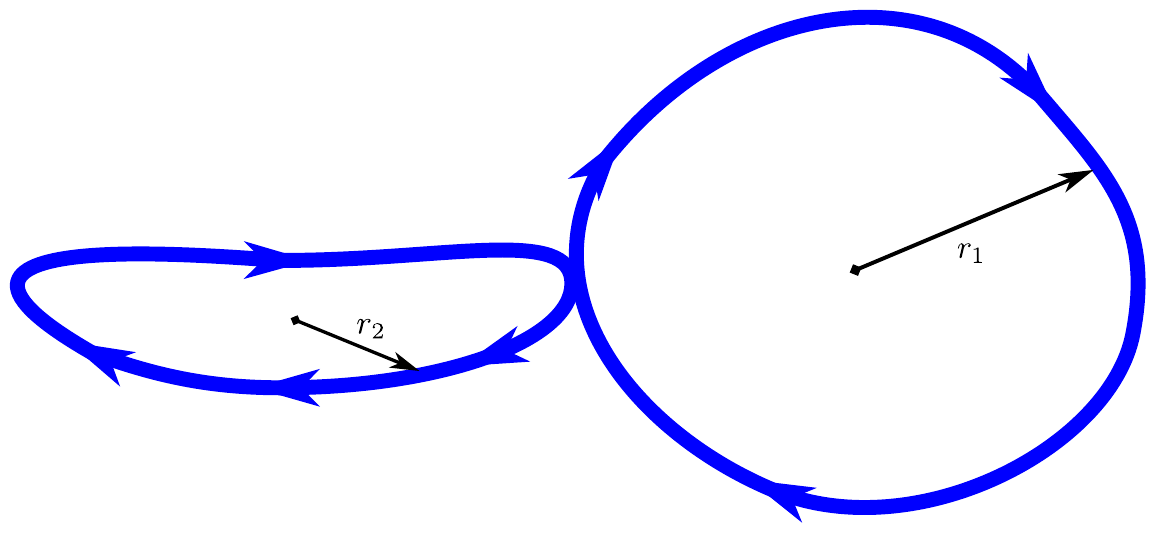}
\caption{A schematic plot of the formation of two axion string loop with radius $r_1$ and $r_2$ from a self crossing event. The surface of the two circular loops are perpendicular to each other so that the mutual inductance is zero.}\label{fig:twoloop}
\end{figure}

The first type of process that can change the total charge on a string is the self-crossing of strings. Even if the charge on a whole string loop - axion system starts off zero, crossing can impart charge on the two resulting string loops (see Fig.~\ref{fig:twoloop}). To see this effect most cleanly, let us consider the case of a massive axion as the electric charge stored in the axion profile outside of the string is confined to the domain wall, where the axion field gradient is non-zero. 
In this situation, an axion string loop is filled in by an axion domain wall.  

During the process shown in Fig.~\ref{fig:twoloop}, the original string with total length $2 \pi (r_1+r_2)$ breaks down into two smaller circular loops each with lengths $2 \pi r_1$ and $2 \pi r_2$, respectively.
Neglecting redistribution of the charge density, the two string loops before crossing carry the same charge density  $\lambda_Q$ and current $I$.  These are related by the relation, given the inductance of a circular loop,
\begin{equation}
2 \pi \lambda_Q (r_1+r_2) + \frac{e^2}{2 \pi} I \left(r_1 \log(\frac{8 r_1}{e^2 r_s})+ r_2 \log(\frac{8 r_2}{e^2 r_s})\right) \approx Q_{\rm tot} = 0 ,
\end{equation}
where $r_s$ is the radius of the string and $e$ in this and the following equation is Euler's number rather than electric charge.
After the crossing event has occurred, the resulting two axion strings carry a non-zero charge
\begin{equation}
\label{eq:chargeseparation}
\left|Q_{\rm tot}^{(1,2)}\right| \approx \frac{2 \pi \lambda_Q r_1 r_2 \log(r_1/r_2)}{r_1 \log(8 r_1 f_a/e^2)+ r_2 \log(8 r_1 f_a/e^2)},
\end{equation}
which is zero only if $r_1 = r_2$.  This shows with an explicit example how string crossings change the total charge of the axion - domain wall system.  

The fact that string crossings can change the total charge of the system is independent of whether the axion has a mass, the exact geometry of the crossing, or if the string itself comes with a significant charge density inhomogeneity. This suggests that the string dynamics in the early universe can easily turn an axion string with charge on the string core and in the axion profile outside the string core (which sums up to zero) into string loops that do carry macroscopic total charge. As a result, it is very conceivable that all strings are somewhat electrically charged.

Another example of string dynamics that lead to charge dissipation is kink formation, which allows for high energy charged particles to be released.  We refer interested readers to Ref.~\cite{Vilenkin:2000jqa} for more details.

\subsection{Charged particle scattering}

The second type of process that can change the total charge on a string is the collision between the bound state particles and the fermions in the string core.  Such a charge loss mechanism starts to be effective whenever the charge density $\lambda_Q$ surpasses roughly the scale $m_{PQ}$. 
As discussed in section~\ref{Sec: gmt}, a charged axion string quickly starts to neutralize its charge using bound state particles.  These particles have a binding energy of order $\lambda_Q \log \left(f_a L \right)$. 
This extremely dense cloud of particles have a significant wave-function overlap with the string core.  
When the string obtains a charge larger than $\lambda_Q \gtrsim m_{PQ}$, scattering of the cloud around the string with the string localized fermions can kick out the zero mode fermion on the string. Such a discharging process is analogous to similar processes well known in the quantum hall community~\cite{Stone:2012ud}.

Once such a process is energetically possible, the discharging of the string becomes efficient. The rate of this discharging process depends on the only energy scale in the problem $\lambda_Q$ up to some logarithmic correction and potentially other small numbers in the problem. Therefore, we expect the string to discharge at a rate that is at least
\begin{equation}
\frac{{\rm d} \lambda_Q}{{\rm d} t} \sim \alpha^2 \lambda_Q^2 .
\end{equation}
in the middle of the ultra-dense plasma around the string of order $\lambda_Q^3$. 
This discharge rate is extremely fast, so that any axion string's charge density can never become larger than the scale $m_{PQ}$.
The only requirement for the above discussion to hold is that there is a light particle with masses much smaller than $m_{PQ}$ charged under the same gauge groups the string localized zero mode fermion is charged under. For QCD axion string, such particles can be the SM quarks charged under QCD, while for an axiverse axion, such particles could be any SM states with electroweak quantum number.

\subsection{Tunneling of zero modes off of the string}

As soon as $\lambda_Q > m_{PQ}$, zero mode PQ fermions can tunnel off of the string.  This is easily seen from the classic quantum mechanics example of a wave packet with energy $\omega$ incident on a barrier of height $V_0$ when the energy on the other side of the barrier is $V'$.  As long as $\omega$ is larger than $V'$, then it can tunnel through the barrier with a probability $\sim e^{-(V_0 - \omega) L}$ where L is the length of the barrier.  This exponential suppression is lost when $\omega \sim V_0$ or $L \sim 1/V_0$.

In the case of axion strings, we have $V_0 \sim f_a$ and $\omega \sim \lambda_Q$ where we are making the simplifying assumption that the mass of the radial mode $\sim f_a$.
Zero mode fermions can leave the string when $\omega \sim \lambda_Q > m_{PQ}$.
They start leaving the string with non exponentially suppressed rates when $\lambda_Q > f_a$ or when $L \lesssim 1/f_a$.  Assuming a radius of curvature $R$, one can estimate $L \sim \sqrt{R/f_a}$ so that $L \sim 1/f_a$ when $R \sim 1/f_a$ and the string can barely even be called a string.
From this estimate, we see that this manner of discharging the string is sub-dominant as compared to scattering with the bound states.

\subsection{String Monopole interactions}

The third type of process that could change the total charge on a string is a string-monopole interaction. It is very well-known that a magnetic monopole that passes through a normal superconducting loop in the lab will lead to the appearance of a small superconducting current on the loop, corresponding to a magnetic flux inside the superconducting loop that is equal to the total magnetic flux that is carried by the magnetic monopole. 
In the case of the axion string, a monopole passing through an axion string loop changes the total charge in the string - axion system, $Q_{\rm tot}$, by a unit quanta~\cite{Naculich:1987ci}.  However, given that we do not see magnetic monopoles anywhere in our universe (see Ref.~\cite{Zyla:2020zbs} and references within), this possibility is of mostly academic interest. As a result, we will not go into details in describing the exact dynamics of such an interaction.

\subsection{Observable consequences}

In the mechanisms discussed above, zero-mode fermions can leave the string.  These PQ fermions can have non-trivial observable consequences if the PQ fermions are stable or meta-stable.  This can easily occur if the PQ fermions carry an electric charge that does not allow it to decay into the lighter standard model particles.  
These stable particles can cause well known cosmological problems if the universe reheats above the mass of the PQ fermions (see e.g. Ref.~\cite{Zyla:2020zbs} and references within for ways in which this problem can be resolved). 

A stable PQ fermion does not necessarily pose a problem in and of itself.  For example, if the PQ fermions are charged under some additional gauge group that confines at a very high energy scale as happens in the case of a composite axion~\cite{Randall:1992ut}.  In the large N limit of these scenarios, the PQ fermions will confine mainly into unstable mesons instead of baryons which allows the PQ fermions to annihilate into light Standard Model states.  These models typically have domain wall problems, but these issues can be mitigated.

When moving through a magnetic field, axion strings can produce many PQ fermions.
As discussed in Sec.~\ref{sec:chargeup}, when a string passes through a galaxy the string can obtain a PQ charge as large as $10^{45}$. If the PQ fermions are light enough, then these PQ fermions are emitted by the string into the galaxy.  
For a typical galaxy, this amounts to a concentration of
\begin{equation}
\frac{n_{\rm CHAMP}}{n_p} \sim 10^{-25},
\end{equation}
which is boarder line accessible with heavy water searches (see Ref.~\cite{Burdin:2014xma} and references within) depending on the mass of the PQ fermion.  Of course, we would need to be incredibly lucky or unlucky for a string to pass though our galaxy in the distant past, since the probability is almost negligible.

The situation can become more interesting if there were large primordial magnetic fields in the early universe, for example as a result of some first order phase transition. The large primordial magnetic fields can give rise to the production of stable and meta-stable remnants starting from the time of the phase transition.
If the magnetic field in the early universe is large enough that the emission of PQ fermions takes all of the kinetic energy of the strings, then these charged remnants can have a density up to
\begin{equation}
\frac{n_{\rm CHAMP}}{n_p} \sim \frac{\xi T_{\rm PT} f_a^2 m_p v_s^2}{T_{\rm eq} m_{\rm pl}^2 m_{PQ}} \approx 10^{-13} \left(\frac{T_{\rm PT}}{{\rm GeV}}\right)\left(\frac{f_a}{10^{14}{ \,\rm GeV}}\right)^2 \left(\frac{10^{14}{ \,\rm GeV}}{m_{PQ}}\right) \left(\frac{\xi}{10}\right) \left(\frac{v_s}{0.1}\right)^2
\end{equation}
where $T_{\rm PT}$ is the temperature at the time of the phase transition. 
If the PQ fermions cannot take away all of the kinetic energy of the string, then the charged remnants have a density
\bea
\frac{n_{\rm CHAMP}}{n_p} \sim \frac{\xi B m_p v_s}{T_{\rm eq} T_{\rm PT} m_{\rm pl} } \approx 10^{-8} \left(\frac{T_{\rm PT}}{{\rm GeV}}\right) \left(\frac{B}{ T_{\rm PT}^2 }\right) \left(\frac{\xi}{10}\right) \left(\frac{v_s}{0.1}\right) .
\eea
If the stable remnants are light, they can be looked for in the various CHAMP searches today, while if these stable remnants are heavy, they can easily make up a significant fraction of the energy density of the universe. For example, if these heavy stable remnants have masses $m_{PQ} \sim f_a$, then they can easily have an energy density that is comparable to the dark matter density today even with a relatively small phase transition temperature.

If the PQ fermions are charged under a dark $U(1)$ and a phase transition happened in a dark sector, such a mechanism could easily be used to populate the universe with very heavy dark matter.  It would be interesting to pursue dark matter models of this sort.

\section{Implications for QCD axion strings}
\label{sec:QCD}

In this section, we briefly discuss the implications of our findings on the QCD axion. It would be very exciting if the effects of anomaly inflow studied in this paper could allow us to produce QCD axion dark matter with much larger axion masses. 
If QCD axion strings obtain charge, then the charge may affect the behavior of the 
axion string.
There are three points where QCD axion strings might be affected : stability of vortons,
string crossing, and deviation from scaling due to friction.
Of these, we find that the first two are not relevant while new sources of friction may be important for other
signatures such as gravitational waves.

\paragraph{Stability of vortons :}

An axion string loop with radius $R$, current $I$, and charge density $\lambda_Q$ might be stabilized at a finite size by the electric repulsion as well as angular momentum conservation\footnote{We thank Robert Lasenby for many helpful conversations.} when the charge density and the current approaches the critical value of $\lambda_Q \sim I \sim f_a$. As a result, the stability of the vortons will be affected if the charge and current on the string can decay. There are two different ways a vorton can lose charge or current. 

The first way a vorton can lose charge is if the PQ quark can decay.  PQ quarks are required to be unstable and decay so as to not overclose the universe due to their production at the PQ phase transition or during the subsequent string evolution.  In particular, charging the PQ quarks under a strong group cannot solve the problem. To see this note that, while during the PQ phase transition it is conceivable that the stable PQ quark will mostly form mesons and decay, the charged strings would shed its charges in PQ quarks that would dominantly form baryons which will remain stable.

If heavy PQ quarks can decay into standard model quarks, the PQ quark zero mode present on small axion string loops will also decay as long as $\lambda_Q > m_{SM}$.  The curvature of the string loop provides momentum non-conservation that allows a PQ quark to decay into standard model quarks.  This suggest microscopic vortons with $Q \lesssim f_a/m_q \sim 10^{13}$ cannot be stable~(see Ref.~\cite{Ibe:2021ctf} for a more quantitative analysis of this effect).  
 
The second way a vorton can lose charge is through scatterings between the PQ charge and the cloud of standard model particles around it as discussed in the previous section. For an axion string carrying $\lambda \sim f_a$, such a density of charged particles is much larger than the ambient density of the plasma, which greatly enhances the rate at which charge is lost.  In particular, the axion string will start to lose charge efficiently once it is possible to scatter PQ quarks off of the string.

An axion string loop with radius $R$ and charge $Q$ in the early universe carries energy mainly in the string tension, angular momentum in the current, and the electric static field around the string
\bea
\label{eq:killer}
E \approx (2\pi R) \left(\pi f_a^2 \log (R/r_s)\right) + \frac{Q^2/e^2}{2 R} + \frac{Q U}{2}, 
\eea
where $U$ is the electric potential on the string core and $r_s \sim 1/f_a$ is the radius of the string core. In the following, let us first neglect the cloud of Standard Model particle bound states and the charge stored inside the axion cloud outside the string core as it would decrease the energy held in the electromagnetic fields and shrink the angular momentum barrier and make the charge $Q$ even less likely to have an effect.  In the case of a naked string, we can find the electric potential $U$ to be
\begin{equation}
U = \frac{\lambda_Q R}{\pi}\frac{\mu}{\sqrt{4 R (R-r_s)}}\int_0^{\pi/2} \frac{1}{\sqrt{1-\mu^2 \sin^2 (\phi)}}\mathrm{d}\phi,
\end{equation}
where
\begin{equation}
\mu^2 = \frac{4 R (R-r_s)}{(2R-r_s)^2}.
\end{equation}
In the limit where $r_s \ll R$, the integral is logarithmically divergent in the ratio of $r_s/R$, reproducing the expectation in section~\ref{sec:kamehamehas}:
\bea
U \approx  \frac{Q}{4 \pi^2 R} \log (\frac{R}{r_s}).
\eea
In the case where there is a single species of zero mode fermion and a moderate size of the log, the angular momentum is usually the main effect that stabilizes the vorton. In general cases, both can be comparably important.


Since the particles inside the electromagnetic field will not be eigenstate of the physical momentum, when $\lambda_Q$ saturates the maximum value of $m_{\rm PQ}^2/(2 U)$, PQ quarks can be scattered off the string core.   This condition can be obtained by requiring that the zero mode scattering obeys $s \geq m_{\rm PQ}^2$.  We find, plugging in this condition into equation~\ref{eq:killer},
\begin{equation}
\frac{\mathrm{d} E }{\mathrm{d} R} \approx 2 \pi^2 \left(f_a^2  (\log (R/r_s)+1) - m_{\rm PQ}^2/4\pi - \frac{\pi}{ e^2 \log (R/r_s)}m_{\rm PQ}^2  \right).
\end{equation}
From this, we see that in order for an axion string loop not to be dominated by its tension, $m_{\rm PQ}^2  \gtrsim  4 \alpha f_a^2 (\log (R f_a))^2$
is required, where the $\log (R f_a)$ is at least of order 30 for $R \gtrsim 1/{\,\rm MeV}$. Similarly, if the electric potential is the dominant force that helps maintain a balance, one would require $m_{\rm PQ}^2  \gtrsim  4 \pi f_a^2 \log (R f_a)$.

As a result, we find that the large charges required to stabilize macroscopic axion strings cannot exist and that macroscopic vortons are not stable in the generic case where the Yukawa is $\mathrm{O}(1)$. If a string loop carries a charge $Q$, then it will radiate axions and shrink in size until $\lambda_Q \sim m_{PQ}, f_a$.  At this point, the string discharges its charge faster than it emits axion radiation so that it shrinks in size while maintaining a roughly constant $\lambda_Q$. PQ quark decay takes over when the axion string shrinks to sizes of order fm to pm. Eventually the string reaches a size of order its radius and vanishes. 
If the Yukawa coupling $m_{\rm PQ}/f_a$ is much larger than $\sim 5$, then it is possible that stable macroscopic vortons may exist.  In this scenario details of the distribution of the charge density outside the ring matters and a more in-depth study is required. 

The discharge through scattering PQ fermions with a bound state particle occur by having a particle in a outer bound state scatter with the PQ fermion and transition to a inner bound state, like photon emission through de-excitation or inverse beta decay in atomic physics. Similar requirements on the strength of the Yukawa coupling can be derived from analysis of bound states.

\paragraph{String crossing :}

The other way in which charge may affect the behavior of QCD axion strings is to change the string scaling behavior by preventing string crossing.  As showed in Sec.~\ref{Sec: gmt}, the axion string is very much neutral to a distant observer. As this close-to-neutral string oscillates, its energy is constantly switching between kinetic and potential energy.  Thus its kinetic energy is of order its total energy.  Comparing kinetic energy to the electromagnetic repulsion, one again finds Eq.~\ref{eq:killer} that leads to the same conclusions as before. 

\paragraph{Friction : }

It is possible that strings do not reach the scaling regime until some time well after they are produced.  The reason for this is that frictional effects can stop the axion string from moving at the near relativistic speeds required to maintain scaling.
As we demonstrated in Eq.~\ref{Eq: photon axion scatter}, the scattering rate of photons off of the axion string
is quite large, $\frac{d \sigma}{d z} \sim \alpha^2/\omega$.  The friction per unit length coming from scattering with the thermal bath is
\bea
f_{\rm friction} = \frac{F}{L} \sim \rho \frac{d \sigma}{d z} v_s v ,
\eea
where $\rho$ is the energy density of the thermal bath doing the scattering, $v_s$ is the speed of the string and $v = 1$ is the velocity of the particles doing the scattering.
Comparing this to the tension, $f_{\rm tension} \sim f_a^2 \log ( \frac{R}{r_s})/L$, we get that the equilibrium velocity of the string is
\bea
v^{\rm eq}_s \sim \frac{f_a^2 \log ( \frac{R}{r_s})}{\alpha^2 T M_{\rm pl}} ,
\eea
which is valid for for string lengths of order Hubble and whenever the equilibrium velocity is less than 1.
As long as $v^{\rm eq}_s \lesssim 0.1$, scaling is not reached.  As soon as $v^{\rm eq}_s \sim 0.1$, scaling is reached within a few e-folds~\footnote{Strings have been simulated with a variety of initial conditions.  Scaling is always reached within a few e-folds.}.  As long as the temperature at which scaling starts is above the GeV scale, then the QCD axion string reaches scaling and the standard results hold to the extent that the scaling regime is insensitive to initial conditions.

The temperature at which the strings reach scaling is
\bea
T_{\rm scaling} \sim \frac{f_a^2 \log ( \frac{R}{r_s})}{\alpha^2 M_{\rm pl}} \sim 10^4 {\rm GeV} \left ( \frac{f_a}{10^8 {\rm GeV}} \right )^2 \gg {\rm GeV} .
\eea
Thus we find that in the allowed region of parameter space, the QCD axion abundance coming from strings is at most logarithmically effected by these frictional type effects.  For $f_a \sim 10^{11}$ GeV, where
the correct DM abundance is reached, it is likely that friction is never important.

An axion string can carry a large current.  This large current can create a magnetic field that scatters incoming particles increasing its effective size~\cite{Chudnovsky:1986hc,Dimopoulos:1997xa}.  Thus an axion string with a large current has a larger scattering length and experiences an increased friction.  
Note that this current is also be shielded by the standard model particles in the bound state surrounding the string.  In a thermal bath with temperature $T$, the string core is coated by a dense cloud of bound states that moves with the same speed and in the same direction as the string localized PQ fermions.

As an estimate, let us take the current on the string to be its maximal value.  
However, since particle production process can also exist in regions where $B\gtrsim E$ and even regions where $E=0$ as long as $\sqrt{B}$ is much larger than the lightest charged particle in the standard model, the electron. Bound state particles likely screen the magnetic field so that it also falls as $B \sim \frac{1}{r^2}$.
If $B \sim T^2$, then the thermal bath of particles will scatter off of the magnetic field and slow down the string.
Thus we find that the largest the effective cross section can be is $\frac{d \sigma}{d z} \sim \alpha/T$.
Following the logic presented earlier, we find that 
\bea
T_{\rm scaling} \sim \frac{f_a^2 \log ( \frac{R}{r_s})}{\alpha M_{\rm pl}} \sim 10^2 {\rm GeV} \left ( \frac{f_a}{10^8 {\rm GeV}} \right )^2 \gg {\rm GeV} .
\eea
That is, given the most relevant parameter spaces of interest~\cite{Gorghetto:2018myk,Gorghetto:2020qws,Buschmann:2019icd} and current constraints on the scale $f_a$ from supernova cooling (see~\cite{Chang:2018rso} and references within), the string network will move back towards scaling at temperatures orders of magnitude larger than the temperature where QCD phase transition occurs. Therefore even assuming a large current on the axion strings, these effects are unlikely to be important in helping to produce very heavy axion dark matter.

To summarize, electromagnetic effects are subdominant to effects coming from the string tension due to bound states formation discussed in~\ref{Sec: gmt}.  The QCD axion string behavior is unfortunately qualitatively unaffected by the charge density or current it may be carrying. 
In the previous discussion, we assumed that the PQ quarks are stable as this
is the scenario where they are most likely to have a large effect.  If the PQ quarks
were unstable, then there are other ways for charge to leave the string, such as 
decay, which may be even more efficient than the ones we considered.

\section{Flux (non) quantization}\label{sec: littleparks}

Charge and flux (non) quantization is one of the most surprising behavior of axion strings, as well as Witten's superconducting strings. In Sec.~\ref{sec:naculich} and Sec.~\ref{sec:galacticB}, we discussed in detail the scattering of photons with both Witten's superconducting string and the axion string. The scattering cross section suggests that most of the magnetic flux passes through the string core unaffected and only a small fraction $\mathcal{O}(\alpha \eta')$ of the magnetic field lines are scattered upstream.  This somewhat surprising result apparently contradicts the common lore that the magnetic flux through a superconducting ring is quantized and constant. 

The non-constant nature of the flux can be understood by realizing that Witten's superconducting string as well as the axion string are both super thin. The thickness of the string is comparable to the penetration depth of the electromagnetic field, which makes the flux crossing process no longer a tunneling event. This allows the magnetic flux inside a superconducting string loop to be no longer conserved.

The second issue is whether the magnetic flux through a superconducting string or axion string is quantized or not. Does the magnetic flux through a superconducting string or axion string always change by integer multiple of the unit flux? Surprisingly, the answer to this question can already be obtained from the Little-Parks experiment done in 1962~\cite{PhysRevLett.9.9} (see Michael Tinkham's book~\cite{Tinkham2004Introduction} for a beautiful discussion). Using a thin superconducting cylinder with thickness comparable to the penetration depth of the electromagnetic field, the Little-Parks experiment demonstrated that fluxoid instead of flux is quantized in a superconducting ring, and it is perfectly fine to have flux non-quantization on the boundary of a superconductor as an intermediate state in the Ginzburg-Landau theory. Similarly, in the case of the 1+1d superconducting string, it can be shown analogously that it is energetically favorable to have flux non-quantization as magnetic field passes through.

The non-conservation and non-quantization of magnetic fluxes are also true in the case of the axion string, which is very similar to the edge of a quantum hall system (see for example Ref.~\cite{Stone:2012ud}). Here, the non-conservation and non-quantization of magnetic fluxes become non-conservation and non-quantization of electric charge on the edge (and hence also the bulk) of a quantum hall system, as the Hall current can bring charges onto the edge of a quantum hall system. Similarly, in the case of the axion string, the charge stored in the axion string core and the axion profile around it are also not separately conserved or quantized due to the anomaly.
While their sum is conserved and quantized, they are not individually conserved or quantized.

\section{Electromagnetic fields around an axion string in the absence of bound states}\label{sec:fun}

In this section, we show how the electric field and magnetic fields behave around an axion string in the absence of bound states.  The equations of motion around a string are
\bea
\vec \nabla \cdot \vec E &=& \rho - \anomaly \alpha \, \vec B \cdot \vec \nabla \frac{a}{f_a} \\
\vec \nabla \times \vec B &=& \vec J - \anomaly \alpha \, \vec E \times \vec \nabla \frac{a}{f_a} 
\eea
To solve these equations, we take the approximation that the current and line charge has a radius $R \sim 1/m_{PQ}$ that is uniformly distributed.
For $r < R$, the axion modified Maxwell's equations can be solved to give
\bea
\vec E &=& \frac{\lambda_\text{eff}}{2 \pi R^2} r \, \hat r  \qquad \lambda_\text{eff} = \frac{2 \lambda_Q - \anomaly \alpha I}{2 - \anomaly^2 \alpha^2/2} \\
\vec B &=& \frac{I_\text{eff}}{2 \pi R^2} r \, \hat \theta \qquad I_\text{eff} = \frac{2 I - \anomaly \alpha \lambda_Q}{2 - \anomaly^2 \alpha^2/2} .
\eea
For $r > R$, the solutions are
\bea
\vec E &=& \frac{\hat r}{4 \pi r}  \left [ \frac{r^{\anomaly \alpha}}{R^{\anomaly \alpha}} \left ( \lambda_\text{eff} - I_\text{eff}  \right ) + \frac{R^{\anomaly \alpha}}{r^{\anomaly \alpha}} \left ( \lambda_\text{eff} + I_\text{eff}  \right ) \right ]  \\
\vec B &=& \frac{\hat \theta}{4 \pi r}  \left [ \frac{r^{\anomaly \alpha}}{R^{\anomaly \alpha}} \left ( I_\text{eff} - \lambda_\text{eff} \right ) + \frac{R^{\anomaly \alpha}}{r^{\anomaly \alpha}} \left ( I_\text{eff} + \lambda_\text{eff} \right ) \right ]  .
\eea
In the case of interest for this article, $I \approx \lambda_Q$, in which case things simplify to
\bea
|E| = |B| = \frac{\lambda_\text{eff}}{2 \pi r} \frac{R^{\anomaly \alpha}}{r^{\anomaly \alpha}} .
\eea
From this, it is clear that we can to leading order in $\anomaly \alpha$ neglect the screening of the electric/magnetic fields of the wire coming from the axion cloud surrounding it.

\bibliographystyle{JHEP}
\bibliography{draft}

\end{document}